\documentclass[aps,prc,twocolumn]{revtex4}
\usepackage{amsmath}
\usepackage{amsfonts}
\usepackage{graphicx}
\def\slash#1{\not\!#1}
\usepackage{color}

\begin{document}
\title{Chiral condensate at finite density using chiral Ward identity}
\author{Soichiro Goda}
\affiliation{Department of Physics, Graduate School of Science, Kyoto University, 
Kyoto 606-8502, Japan}
\email{gouda@ruby.scphys.kyoto-u.ac.jp}
\author{Daisuke Jido}
\affiliation{Department of Physics 
Tokyo Metropolitan University 
Hachioji, Tokyo 192-0397, Japan}


\begin{abstract}
In order to study partial restoration of the chiral symmetry at finite density, we investigate the density corrections of the chiral condensate up to the next-to-leading order of density expansion using the chiral Ward identity and an in-medium chiral perturbation theory.
In our study, we assume that all the in-vacuum quantities for the pion, the nucleon and the $\pi N$ interaction are determined in vacuum and focus on density expansion of the in-medium physical quantities.
We perform diagrammatic analysis of the correlation functions which provide the in-medium 
chiral condensate. This density expansion scheme
shows that the medium effect to the chiral condensate beyond the linear density comes from 
density corrections to the $\pi N$ sigma term as a result of the interactions between pion and nucleon
in nuclear matter.  
We also discuss that higher density contributions beyond order of $\rho^{2}$ cannot be 
fixed only by the in-vacuum $\pi N$ dynamics and we encounter divergence in the calculation of the $\rho^{2}$ order corrections of the chiral 
condensate.
To remove the divergence, we need NN two-body contact interaction, which can be fixed in vacuum.
\end{abstract}

\pacs{11.30.Rd, 21.65.Jk, 11.30.Rd, 12.39.Fe}
\maketitle

\section{Introduction}

Dynamical breaking of chiral symmetry (DB$\chi$S) is one of the important phenomena of QCD for low-energy hadron spectrum and dynamics of light hadrons. The light pseudoscalar mesons, $\pi$, $K$ and $\eta$, are identified as the Nambu-Goldstone bosons of DB$\chi$S, and the quark mass generation  is also explained by DB$\chi$S. The quark condensate $ \langle \bar{q} q \rangle $ is one of the order parameters of DB$\chi$S and its magnitude characterizes the QCD vacuum. Since DB$\chi$S is a phase transition phenomenon, such dynamically broken symmetry is expected to be restored in extreme environment, such as high temperature and/or high baryonic density. It is very significant to confirm phenomenologically that DB$\chi$S really takes place in the QCD vacuum. 

One of the proofs of DB$\chi$S is to make sure of partial restoration of chiral symmetry in nuclear matter.
The partial restoration of chiral symmetry is an incomplete restoration of chiral symmetry with sufficient reduction of the magnitude of the quark condensate. Recent observations of pionic atom spectra, especially precise measurement of the isotope dependence of deeply bound pionic atoms~\cite{Su}, and low energy pion-nucleus scattering~\cite{Friedman:2004jh,Friedman:2005pt} have found out that the $b_{1}$ parameter appearing in the pion optical potential
is substantially enhanced in nuclei. With this fact and theoretical examination~\cite{KKW,JHK}, it turns out that the magnitude of the quark condensate does decrease about 30\% at the saturation density. 

The reduction of the quark condensate in nuclear medium also leads to various phenomena, for instance, attractive enhancement of scalar-isoscalar $\pi\pi$ correlation in nuclei~\cite{Hatsuda:1999kd,Jido:2000bw,Hyodo:2010jp} and the suppression of the spectrum difference between the chiral partners, such as $\rho$-$a_{1}$~\cite{Weinberg:1967kj,Kapusta:1993hq} and $N$-$N(1535)$~\cite{Detar:1988kn,Kim:1998up,Jido:1998av,Jido:2001nt}. The experimental observations of these phenomena can be further confirmation of partial restoration of chiral symmetry in nuclear medium.  For instance, one could observe the reduction of the $N$-$N(1535)$ mass difference from the formation spectrum of the $\eta$ mesonic nuclei~\cite{Jido:2002yb,Nagahiro:2003iv,Jido:2008ng}. The mass difference between the $\eta$ and $\eta^{\prime}$ mesons is also responsible for the quark condensate through the $U_{A}(1)$ anomaly effect~\cite{Lee:1996zy,Jido:2011pq}.

These phenomena are caused by substantial quark dynamics, but because we have quark-hadron duality in the description of hadron dynamics, these phenomena should be also described in terms of hadron dynamics, such as nuclear many-body theories. This means that if one could describe the suppression of the spectrum difference of the chiral partners in a  nuclear many-body theory, this does {\it not} rule out partial restoration of chiral symmetry. To a greater extent, once one could understand hadronic phenomena in terms of quark-gluon dynamics, one would have more substantial and deeper insight of hadron dynamics, which will bring us its more systematic understanding in terms of QCD. 

Theoretically the reduction of the quark condensate in the nuclear medium is naturally expected according to the model independent low density relation~\cite{DL}, in which the ratio of the in-medium and in-vacuum quark condensates is given by the $\pi N$ sigma term together with the in-vacuum pion mass and pion decay constant. This relation is derived under the linear density approximation. The sign of the sigma term determines the fate of the in-medium quark condensate. Since the sigma term extracted from $\pi N$ scattering data has a positive sign, the quark condensate should be reduced, at least, in low density limit. However, one does not know up to which density one can apply the linear density approximation. For further detailed understanding, one needs calculation beyond the linear density based on effective theories. 

Because the quark condensate is not a direct observable in experiments, one needs theoretical examination to conclude partial restoration of chiral symmetry phenomenologically. In Ref.~\cite{JHK}, an exact sum rule which relates the quark condensate and hadronic observables has been derived by using the chiral Ward indentity. In the linear density approximation, the in-medium quark condensate can be written in terms of the in-medium temporal pion decay constant and the pion wavefunction renormalization constant. The importance of the wavefunction renormalization in in-medium chiral effective theories has been also discussed in Refs.~\cite{KKW,Jido:2000bw}. With this relation, the reduction of the quark condensate has been phenomenologically confirmed by using the in-medium pion decay constant extracted from pion-nucleus dynamics~\cite{Su} and the wavefunction renormalization constant extracted from pion-nucleon scattering~\cite{JHK}. Since this proof of the partial restoration of chiral symmetry in the nuclear medium is based on the linear density approximation of in-medium quantities, precise determination of the density dependence of the quark condensate both in theory and experiment~\cite{KHW,Ikeno:2011mv} is strongly desired.

The in-medium quark condensate has been discussed in various approaches. The in-medium correction of the condensate is given by the pion-nucleon sigma term model-independently at the first order in nucleon density \cite{DL,Coh91} and higher orders are evaluated with mean field calculations using the Nambu-Jona-Lasinio and Gell-Mann-Levy models\cite{Coh91} and the relativistic Brueckner approach\cite{Guo94,Bro96}. In Ref.~\cite{KHW}, the in-medium condensate has been obtained beyond the linear density based on the Hellman-Feynman theorem, in which they have calculated the energy density in nuclear matter based on a chiral effective theory and taken its derivative with respect to the quark mass to obtain the quark condensate. The pion self-energies in asymmetric nuclear matter was calculated based on chiral perturbation theory in Ref. \cite{Kaiser:2001bx}. Reference \cite{Kaiser:2001jx} developed systematic framework of chiral perturbation theory in nuclear matter and calculated the equation of state of isospin-symmetric nuclear matter. In Ref. \cite{Girlanda:2003cq}, a novel formulation of chiral perturbation theory in a nuclear background was proposed and the self-energy and the nuclear optical potential of the charged pion were calculated. In Ref.~\cite{Doring:2007qi} hadronic quantities, such as pion optical potential, have been calculated beyond the linear density.

The goal of this paper is to examine higher density correction beyond linear density and to show a systematic way to calculate the in-medium quantities based on chiral effective theory. For this purpose, we use the formulation proposed in Ref.~\cite{O} and developed in Ref.~\cite{MOW}. In this formalism, one calculates matrix elements in the free Fermi nuclear matter, which are defined by the path integral under the action of the system.  All the interaction between nucleons in matter and pions are assumed to be described in the interaction Lagrangian.In this formulation, one can make a double expansion in terms of Fermi see insertion and chiral order counting. Thus, the expansion scheme is clear.

In this paper we calculate the Ward identity, which connect the quark condensate and hadronic quantities,
based on this in-medium chiral perturbation formulation. This paper is organized as follows.
We explain the chiral Ward identity which relates the chiral condensate with the  correlation function of the chiral currents in section 2.
In section 3, we introduce the in-medium chiral perturbation theory and the in-medium chiral counting scheme.
In section 4, we show the result of the calculation of the in-medium chiral condensate $\langle \bar qq \rangle^{*}$, and finally we devote Sec.5 to the conclusion of the present paper.


\section{Chiral Ward identity}

In order to calculate the density dependence of the in-medium quark condensate, we take the chiral Ward identity approach proposed by Ref.~\cite{JHK}. In this approach, we consider the correlation function of the axial vector current $A^a_{\mu}(x)$ and the pseudoscalar density $P^a(x)$:
  \begin{equation}
   \Pi^{ab}_5 (q)= \int d^4 x e^{iq \cdot x} \partial^{\mu} \langle \Omega | T A^a_{\mu}(x) P^b(0) | \Omega \rangle
  \end{equation}
where $| \Omega \rangle $ is the nuclear matter ground state normalized as $ \langle \Omega | \Omega \rangle = 1$ and is characterized by the proton and neutron densities, $ \rho_p$ and $\rho_n $.
The axial vector current $A^a_{\mu}(x)$ is associated with the SU(2) chiral transformation whose generators are given by $Q_{5}^{a} = \int d^{3} x A^{a}_{0}(x)$.
The pseudoscalar density $P^a(x)$ is defined in terms of the quark field by $P^{a}(x) \equiv \bar q i \gamma_{5} \tau^{a} q(x)$ with the Pauli matrix $\tau^{a}$ for the isospin space and transforms under the SU(2) chiral transformation as $[Q_5^a , P^b (x)] = -i \delta^{ab} \bar{q}{q} (x) $.

Using the operator identity $\partial^{\mu} [TA^a_{\mu}(x) P^b(0)] = \delta(x_{0}) [A_{0}^{a}(x),P^{b}(0)] + T[\partial^{\mu} A_{\mu}^{a}(x) P^{b}(0)]$ and performing the integral in the soft limit $q_{\mu} \to 0$, we obtain the in-medium quark condensate as
\begin{equation}
   -i \delta^{ab} \langle \bar{u}u +\bar{d} d  \rangle^* = \Pi^{ab}_5 (0) - m_q D^{ab}(0) 
   \label{ward}
\end{equation}
where we have written the expectation value $\langle \Omega | \mathcal{O} | \Omega \rangle $ as $\langle \mathcal{O} \rangle^* $ for operator $\mathcal{O}$, and $\Pi^{ab}_{5}(0)$ and $D^{ab}(0)$ are defined as
  \begin{eqnarray} 
&&  \Pi^{ab}_5 (0) \equiv \lim_{q_{\mu} \to 0}  -i q^{\mu} \int d^4 x e^{iq \cdot x} \langle  A^a_{\mu}(x) P^b(0) \rangle^*  \label{eq:Pidef}  \\
&&   D^{ab}(0) \equiv  \lim_{q_{\mu} \to 0} \int d^4 x e^{iq \cdot x} \langle P^a(x) P^b(0) \rangle^*   \label{eq:Ddef}
  \end{eqnarray}
Here we have used the PCAC relation $ \partial^{\mu} A^a_{\mu} = m_q P^a $ with the quark mass $m_{q}$.  Equation~\eqref{ward} implies that  the in-medium quark condensate is written in terms of the Green functions in the soft limit.

We can evaluate the in-medium chiral condensate by calculating this correlation functions $\Pi^{ab}_5 (q)$ and  $D^{ab}(q)$ in the soft limit $q \to 0$.
Up to the next-to-leading order, we will confirm that $\Pi^{ab}_5 (0)$ vanishes off the chiral limit in the soft limit when there are no massless pionic modes which coupled to the axial current $A^a_{\mu}(x)$.
In the chiral limit, the quark condensate can be calculated by   
\begin{equation}
     \delta^{ab} \langle \bar{u}u +\bar{d} d  \rangle^* = \lim_{q_{\mu} \to 0} q^{\mu} \int d^4 x e^{iq \cdot x} \langle  A^a_{\mu}(x) P^b(0) \rangle^* 
  \end{equation}
as discussed in Ref.~\cite{JHK}.
We note that both methods are equivalent when one calculate $ \langle \bar{q}q \rangle$ in the chiral limit.


\section{In-medium chiral perturbation theory}

Chiral effective theories are powerful theoretical tools to describe hadron dynamics based on chiral symmetry and its spontaneous breaking\cite{W,GL1,GL2,GSS}. In this work, we use the in-medium extension of the chiral perturbation theory developed by Refs.~\cite{O,MOW}. In this method, first one defines the generating functional of the correlation functions by taking non-interacting Fermi gas of nucleons as the asymptotic state and assumes all the interaction between nucleons and other internal fields are described by the chiral effective Lagrangian. The in-medium correlation functions are calculated by taking functional derivatives of the generating functional.

Let us consider the non-interacting nucleon system at asymptotic times $t\to \pm \infty$, $ | \Omega_{\rm out} \rangle$ and $| \Omega_{\rm in} \rangle $ as usual scattering theory. Here we assume the unpolarized nuclear matter for simplicity. The in- and out-states are described in terms of the nucleon creation operators $a^\dag ({\bf p}_n)$ with the nucleon momentum ${\bf p}_n$ as 
\begin{displaymath}
 | \Omega_{\rm in,out} \rangle \equiv \prod_n^N a^{\dag} ({\bf p}_n)| 0 \rangle
\end{displaymath}
where the nucleon Fermi gas states are occupied up to the Fermi momentum $k_F$. The proton and neutron densities are given by the Fermi momenta $k_{F}^{(p,n)}$ as
\begin{equation}
   \rho^{i} =  \frac{1}{3\pi^{2}} k_{F}^{i3},
\end{equation}
for $i=p,n$.

The generating functional is given by 
\begin{equation}
Z[J, \eta, \eta^\dag ] = e^{i W[J, \eta, \eta^\dag ]} =  \langle \Omega_{\rm out} | \Omega_{\rm in} \rangle_{J, \eta, \eta^\dag} 
\label{generating}
\end{equation}
under the presence of the external fields $J=(s,p,v,a)$, $\eta$ and $\eta^{\prime}$.
Here, $s,p,v,a$ represent the scalar, pseudo-scalar, vector and axial-vector sources respectively and $ \eta, \eta^\dag$ are the nucleon external sources.
We also define the generating functional for the connected Green functions $W[J, \eta, \eta^\dag]$ in Eq.(\ref{generating}).
The path integral is to be performed for the fields in the Lagrangian, such as the chiral field $U$ and the nucleon field $N$:
\begin{eqnarray}
  Z[J,\eta,\eta^{\dagger}] &=& \int DU DN DN^\dagger \langle \Omega_{\rm out}| N (+\infty) \rangle 
  \nonumber \\ && \hspace{-2em} \times
  e^{i \int dx ({\cal L}_{\pi}+ {\cal L}_{\pi N} + \eta^\dag N + N^\dag \eta )} \langle N(-\infty)| \Omega_{\rm in} \rangle, 
\end{eqnarray}
where $\mathcal{L}_{\pi}$ is the pion chiral Lagrangian and 
we take the $\pi$-$N$ chiral Lagrangian with the nucleon bilinear interaction  $A$ given by the $\pi N$ chiral Lagrangian
${\mathcal L}_{\rm \pi N} = \bar{N} ( i \gamma^\mu \partial_\mu  - m_N - A ) N $.
The operator $A$ is written by the pion fields and its derivatives together with the external fields and is subject to chiral order counting, and $ N = (p,n)^T  $ is the nucleon field with $p,n$ for proton and neutron.
In appendix A, the detailed expression of $A$ is summarized.
The parameters of the Lagrangian are to be fixed in vacuum. 

\begin{figure}[t]
    \centering
    \includegraphics[width=8.6cm]{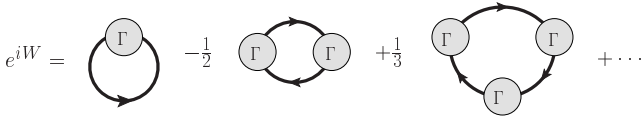}
    \caption{Schematic diagram for the generating functional~\eqref{eq:Zexp} in the expansion of the Fermi sea insertion. The thick line denotes the nucleon propagation in the Fermi sea and $\Gamma$ is the nonlocal vertex given by the in-vacuum $\pi N$ interaction.}
    \label{Z}
  \end{figure}

The integral in terms of the nucleon field can be done easily by using the Gauss integral formula, if the Lagrangian has the bilinear form for the nucleon interaction. As shown in Ref.~\cite{O}, the generating functional is characterized by double expansion of Fermi sea insertions and chiral orders. The Fermi sea insertion is seen as
\begin{widetext}
\begin{eqnarray}
Z[J]
&=& \int DU \exp \Big{\{} i\int dx \Big{[}\mathcal{L}_{\pi \pi} 
- \int \frac{d \bf{p}}{(2\pi)^3 2E(p)} 
  {\rm F.T.\,}
{\rm Tr} \Big{(} i\Gamma(x,y) (\slash{p} + m_N) n(p) \Big{)} 
\nonumber \\  && 
- \frac{i}{2} \int \frac{d \bf{p}}{(2\pi)^3 2E(p)} \frac{d \bf{q}}{(2\pi)^3 2E(q)} 
 {\rm F.T.\, }
 {\rm Tr} \Big{(} i\Gamma(x,x') (\slash{q} + m_N) 
 n(q) i\Gamma(y',y) (\slash{p} + m_N) n(p) \Big{)}
  + \cdots  
\Big{]} \Big{\}} \label{eq:Zexp}
\end{eqnarray}
\end{widetext}
where F.T.\ denotes Fourier transformation of the spacial variables except $x$, 
$ E({\bf p})$ is the relativistic nucleon energy $ E({\bf p}) = \sqrt{ {\bf p}^2 + m_N^2 }$
and the nonlocal vertex $\Gamma(x,y)$ is defined by $\Gamma \equiv -iA[1_4-D_0^{-1}A]^{-1}$ , which is given only by the in-vacuum interactions $A$
and the free nucleon propagator $D_{0}^{-1}$. 
The matrix $n(p)$ in the isodoublet space is defined to restrict the momentum integral up to the Fermi momentum  as
\begin{equation}
 n(p) = \left( 
       \begin{array}{cc}
        \theta ( k_F^{p} - | {\bf p} |) & 0  \\
           0        & \theta ( k_F^{n} - | {\bf p} |)   \\ 
       \end{array}         \right) .
\end{equation}
Fig.~\ref{Z} shows the diagrammatic structure of the Fermi sea insertion of the generating functional~\eqref{eq:Zexp}. In the figure, the thick line represents nucleon propagation in the Fermi sea. The chiral expansion is given by the expansion of the nonlocal vacuum vertices
\begin{equation}
i \Gamma = A + A D_{0}^{-1} A + A D_{0}^{-1} A D_{0}^{-1} A+ \cdots \label{eq:Gamma}
\end{equation}
together with the chiral expansion for the bilinear local vertex $A$.  
Using the generating functional~\eqref{eq:Zexp}, we can define the in-medium pion Lagrangian $ \tilde{\mathcal L}_{\pi \pi}$ as $Z[J]= \exp [{i \int d^{4}x  \tilde{\mathcal L}_{\pi \pi}}]$.

The connected $n$-point Green functions can be evaluated by taking functional derivatives of $ i W [J] $ defined in eq.(\ref{generating})  with respect to the external sources $J^i$:
\begin{equation}
\langle \Omega_{\rm out} |T {\cal O}_{1} \cdots {\cal O}_{n}  | \Omega_{\rm in} \rangle = (-i)^n \frac{\delta}{\delta J_{1} } \cdots \frac{\delta}{\delta J_{n} } i W[J] ,
\label{Green}
\end{equation}
where ${\cal O}_{i}$ is the corresponding current operator to the external source $J_{i}$.
The current operator $ {\cal O}_{i} $ can be represented in terms of the corresponding quark current, such as the pseudo-scalar current $ P^i = \bar{q} \gamma_5 \tau^i q $ and the axial-vector current $A^i_\mu = \bar{q} \gamma_5 \gamma_\mu \tau^i q $.

The in-vacuum chiral perturbation theory has the chiral expansion scheme in which the pion energy momentum and the small quark mass are counted as small quantities. In addition to these quantities, in the in-medium chiral perturbation theory the Fermi momentum is also regarded as a small quantity, since the Fermi momentum at the normal nuclear density $k_{F} = 270$ MeV is as small as $2 m_{\pi}$.
According to Ref.~\cite{MOW}, chiral order $\nu$ for a specific diagram is given by the following:
  \begin{eqnarray}
&& \nu = 4 L_{\pi} - 2 I_{\pi} + \sum_{i=1}^{V_{\pi}} d_i + \sum_{i=1}^{V_{\rho}} d_{\rho i} \geq 4 \label{counting} \\
&& d_{\rho} = 3n + \sum_{i=1}^n \nu_{\Gamma_i} -4(n-1) ,
  \end{eqnarray}
where $L_{\pi} $ is the number of pion loops, $I_{\pi} $ is the number of the pion propagators, $d_i $ is the chiral dimension coming from the pion chiral Lagrangian ${\mathcal L}_{\pi \pi}$, $d_{\rho}$ is the chiral dimension of the nonlocal in-medium vertex with $n$ Fermi sea insertions and $\nu_{\Gamma}$ is the chiral dimension of the $\Gamma$ vertex.
This counting rule is called standard case in Ref.~\cite{MOW} in which the nucleon propagator is counted as $O(p^{-1})$.
We note that $\nu$ is larger than 4, so that leading order contribution to the in-medium chiral condensate appears from $O(p^4)$.
In this formalism, we can calculate any processes in which pions interact with Fermi gas using this method.
If one follows strictly the chiral expansion scheme, one has to expand also in-vacuum terms and renormalize them order by order. This kind of the expansion is useful in theoretical consideration, while it is not convenient in practical use because expanded quantities are not direct observables. Here we consider that all the in-vacuum quantities for the $\pi N$ dynamics are already fixed by experiments. This implies that the renormalization procedure for in-vacuum values are already done and we do not have to evaluate in-vacuum loop diagrams. 
This is also consistent with having taken $\det (D_{0}-A)=1$ to obtain Eq.~\eqref{eq:Zexp}.

\begin{figure}[t]
    \centering
    \includegraphics[width=8.6cm]{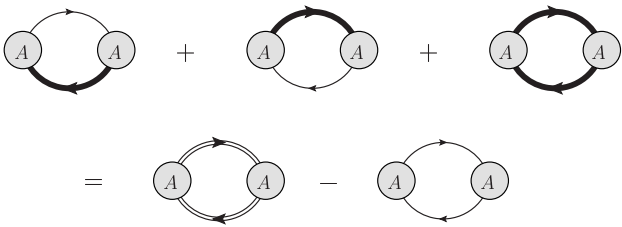}
    \caption{A one-nucleon loop diagram in the nuclear medium. The thick and thin lines represent the nucleon propagations in the Fermi sea and in free space, respectively. The double line denotes the nucleon propagator $G(p)$ given in Eq.~\eqref{Gprop}.}
    \label{example}
  \end{figure}

The expansion scheme of the generating functional in terms of the Fermi sea insertion given in Eq.~\eqref{eq:Zexp} is equivalent to the conventional nuclear many body calculation using the Pauli-blocked nucleon propagator in the Fermi gas
  \begin{eqnarray}
i G (p) &=& iD_{0}^{-1}(p) + i D_m^{-1} (p) . \label{Gprop} \\
iD_{0}^{-1}(p) &=& \frac{i (\slash{p} + m_N)}{p^2 -m_N^2 + i \epsilon} \\
i D_m^{-1} (p) &=& - 2 \pi (\slash{p} + m_N) \delta (p^2 - m_N^2) \theta (p_0) n(p) 
  \end{eqnarray}
To see the equivalence, we examine a one-nucleon loop diagram with two interaction operator $A$ (see Fig.~\ref{example}).
In the conventional approach, this diagram can be calculated using the Pauli-blocked nucleon propagator $G$ by
\begin{displaymath}
 \frac{i}{2} \int \frac{d^4 p}{(2\pi)^4} \frac{d^4 q}{(2\pi)^4} {\rm Tr} \Big{[} (-iA(q-p)) iG(q) (-iA(p-q)) iG(p) \Big{]}.
\end{displaymath}
Using Eq.~\eqref{Gprop}, this can be written as
\begin{widetext}
\begin{eqnarray}
 &&   \frac{i}{2} \int \frac{d^4 p}{(2\pi)^4} \frac{d^4 q}{(2\pi)^4} {\rm Tr} \Big{[} (-iA) iD_{0}^{-1}(q) (-iA) iD^{-1}_{0}(p) \Big{]}
    +  i \int \frac{d^3 {\bf p}}{(2\pi)^3 2E(p)} \frac{d^4 q}{(2\pi)^4} {\rm Tr} \Big{[} (-iA) iD_{0}^{-1}(q) (-iA) i(\slash{p} + m_{N}) n(p) \Big{]}
 \nonumber \\ &&
    +  \frac{i}{2} \int \frac{d^3 {\bf p}}{(2\pi)^3 2E(p)} \frac{d^3 {\bf q}}{(2\pi)^3 2 E(q)} {\rm Tr} \Big{[} (-iA) i(\slash{q} + m_{N}) n(q) (-iA) i(\slash{p} + m_{N}) n(p) \Big{]} \label{eq:2point}
\end{eqnarray}

\end{widetext}
Here we have integrated out in terms of  $p_{0}$ for $D_{m}^{-1}(p)$:
\begin{equation}
  \int \frac{d^{4}p}{(2\pi)^{4}} 2\pi \delta(p^{2} - m_{N}^{2}) \theta(p_{0}) n(p)
  = \int \frac{d^{3} {\bf p}}{(2\pi)^{3} 2E(p)} n(p).
\end{equation}
The first term of Eq.~\eqref{eq:2point} is the nucleon loop in vacuum and should be renormalized into the in-vacuum Lagrangian. The second term can be obtained from the one Fermi sea insertion as appearing in the second term of the argument of exp in Eq.~\eqref{eq:Zexp} after replacing the non-local vertex $\Gamma$ to $AD_{0}^{-1}A$ which is the second term of the chiral expansion of $\Gamma$ in Eq.~\eqref{eq:Gamma}. 
The third term can be obtained in the two Fermi sea insertion by replacing $\Gamma$ to $A$ which is the first term of the chiral expansion. 
In the same way, one can show that Eq.~\eqref{eq:Zexp} contains all the terms of one nucleon loop diagram with three interaction operators given by the conventional approach as
\begin{eqnarray*}
&& \frac{i}{3} \int \frac{d^4 p}{(2\pi)^4} \frac{d^4 q}{(2\pi)^4} \frac{d^4 k}{(2\pi)^4} {\rm Tr} \Big{[} (-iA(q-p)) iG(q) \\ 
&& \hspace{3em} \times (-iA(k-q)) iG(k) (-iA(p-k)) iG(p) \Big{]}
\end{eqnarray*}
with the correct factor except the free nucleon loop. 
Therefore calculation with the in-medium nucleon propagor $G$ is equivalent to use the expansion scheme of the generating functional given in Eq.~\eqref{eq:Zexp}.

\section{Results}

In order to evaluate the in-medium condensate $ \langle \bar{q}q \rangle^* $ with Eq.~(\ref{ward}), we calculate the current-current correlation functions in the soft limit, $ \Pi^{ab}_5 (0) $ and $D^{ab}(0) $ defined in Eqs.~\eqref{eq:Pidef} and \eqref{eq:Ddef}, by using the in-medium chiral perturbation theory.
From eq.~(\ref{Green}), $D^{ab}(0) $  is expressed by the generating functional $W[J] $.
\begin{eqnarray}
D^{ab}(0) &=& \lim_{q \to 0} \int d^4 x e^{iqx} \langle \Omega_{\rm out} | T P^a (x) P^b (0) | \Omega_{\rm in} \rangle \\
&=& \lim_{q \to 0} \int d^4 x e^{iqx} (-i)^2 \frac{\delta}{\delta p^a (x)} \frac{\delta}{\delta p^b (0) } i W[J]. \ \ \ \ \ 
\end{eqnarray}
Here, $P^a (x) , p^a (x)$ are the pseudo-scalar density and the corresponding external field.
Similarly, $ \Pi^{ab}_5 (0) $ is expressed in terms of the generating functional $W[J]$.

In Sect.~\ref{Dfunction}, we list up the Feynman graphs for the calculation of $D^{ab} (0) $  based on the density order counting and evaluate it up to the NLO corrections.
We also list up the Feynman diagram for $ \Pi^{ab}_5 (0)$ evaluate it in Sect.~\ref{cancellation}. We will find that $ \Pi^{ab}_5 (0)$ vanishes off the chiral limit within the NLO corrections by taking the soft limit.
In Sect.~\ref{densitydependence}, we show the density dependence of the chiral condensate within the NLO corrections.
In Sect.~\ref{contactNN}, we discuss higher order corrections beyond NLO. We will find that some diagrams are divergent and show the necessity of the $NN$ contact terms to renormalize the higher order corrections. 
In the following, we consider the symmetric nuclear matter for simplicity.

  \begin{figure}[h]
    \centering
    \includegraphics[width=8.6cm]{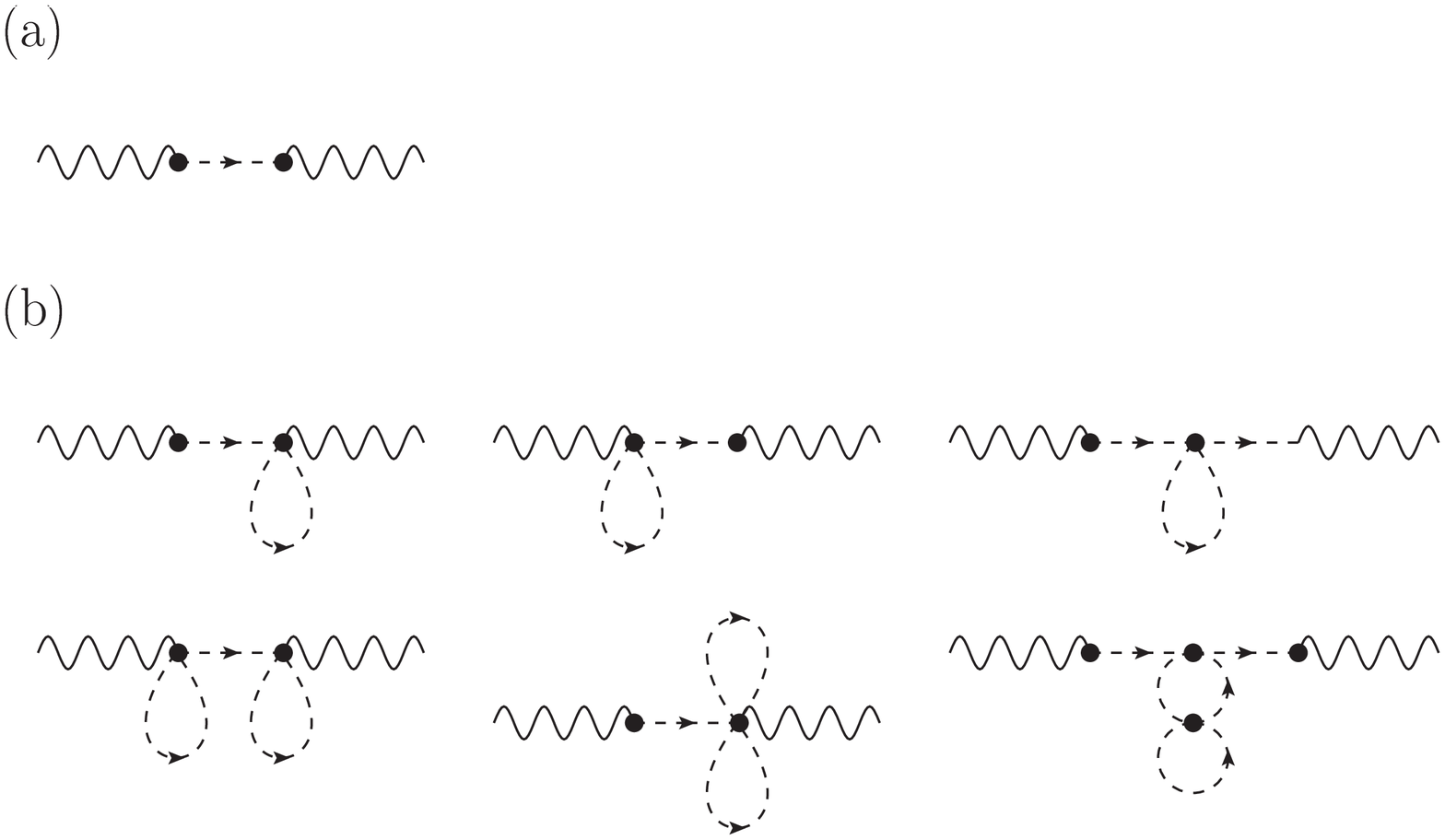}
    \caption{
    Feynman diagrams contributing to the $D^{ab}(0) $ correlation function in vacuum.
    The wavy line denotes the pseudo-scalar density
and the filled circle represents the tree vertex for the pions.
The dashed line stands for the pion propagation.
    The diagram in (a) is for the tree level and the diagrams shown in (b) are
    examples of the radiative correction.}
    \label{p2}
  \end{figure}

\subsection{Calculation of $D^{ab}(0) $}
\label{Dfunction}

We calculate $D^{ab}(0) $ in the soft limit with the finite quark mass. In the following  $\langle P^a (x) P^b (0) \rangle^*$ denotes
the in-medium expectation value  $\langle \Omega_{\rm out} | P^a (x) P^b (0) | \Omega_{\rm in} \rangle$.

First of all, let us evaluate $D^{ab}(0) $ in the vacuum using the in-vacuum chiral 
Lagrangian $\mathcal{L}_\pi$.  We draw the Feynman diagrams for the $\langle P^a (x) P^b (0) \rangle $ correlation function in Fig.~\ref{p2} based on the chiral counting scheme.
In the diagrams the wavy line denotes the pseudo-scalar density
and the filled circle represents the tree vertex for the pions.
Feynman graph (a) shows the leading order graph counted as $O(p^2)$ in  
the chiral counting, while the diagrams (b) are the radiative corrections in the higher
order of the chiral expansion. Calculating $\langle P^a (x) P^b (0) \rangle $ 
in vacuum and taking the soft limit $q \to 0$ in the momentum space, 
we obtain $D^{ab}(0)$ and the in-vacuum chiral condensate through Eq.~(\ref{ward}):
\begin{eqnarray}
D^{ab}_0 (0) &=& (-i)^2 \delta^{ab} \lim _{q \to 0}  (2ifB_0) iD_\pi (q)  (2ifB_0) + \cdots \nonumber \\
&=& - 4 i f^2 B_0^2 \frac{1}{m_\pi^2} \delta^{ab} + \cdots,
\label{vacuum}
\end{eqnarray}
where $f$ is the pion decay constant in the chiral limit, $ B_0$ is one of the low energy constants in the chiral Lagrangian $\mathcal{L}_{\pi}^{(2)}$ given in Appendix \ref{app} , $i D_\pi (q) = i ( q^2 - m_\pi^2 +i \epsilon )^{-1}$ is the free pion propagator and  $ m_\pi $ is the pion mass.
The first term is the leading order contribution of the in-vacuum chiral condensate 
and the dots mean higher order loop corrections, such as those given as the diagrams in Fig.~\ref{p2} (b).
Using the relation between the pion and quark masses in the chiral perturbation theory, $  m_\pi^2 = 2 m_q B_0 + \cdots $, 
we obtain the in-vacuum chiral condensate $ \langle \bar{u}u + \bar{d}d \rangle_0 $:
\begin{equation}
m_q D^{ab}_0 (0) = i(-2f^2 B_0 + \cdots ) = i \delta^{ab}  \langle \bar{u}u + \bar{d}d \rangle_0 .
\end{equation}
In this work we presume that the in-vacuum quantities appearing in the Lagrangian 
are already fixed by the experimental data. 

  \begin{figure}[t]
    \centering
    \includegraphics[width=8.6cm]{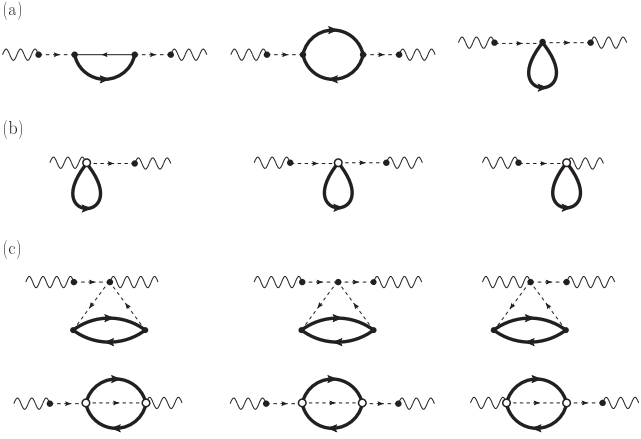}
    \caption{
    Feynman diagrams  contributing to the in-medium chiral condensate.
    The wavy line denotes the pseudo-scalar density,
    the solid line is the free nucleon propagator appearing as the first term of Eq.~(\ref{Gprop}) and the thick solid line denotes the Fermi sea insertion which is the second term in eq.~(\ref{Gprop}) for the Pauli blocking effect. 
  The filled circle represents the vertex in the leading order,  while 
   the unfilled circle is the vertex 
    coming from the in-vacuum subleading $\pi N$ chiral Lagrangian $\mathcal{L}_{\pi N}^{(2)}$.
 (a) The Feynman diagrams for  $O(p^4)$. (b) The Feynman diagrams for the linear density order $O( \rho )$. 
    (c) The Feynman diagrams for the next to leading order in density $O( \rho^{4/3} )$. Three diagrams at the bottom give the double scattering term.}
    \label{density}
  \end{figure}

Second, we consider  the density corrections. According to the chiral counting in Eq.~(\ref{counting}), the leading order contributions start from $ O(p^4) $.
In Fig.~\ref{density} (a), we show all the $ O(p^4) $ diagrams.
In these graphs, the thin solid line is the free nucleon propagator shown as the first term of Eq.~(\ref{Gprop}) and the thick solid line is Fermi sea insertion shown as the second term in Eq.~(\ref{Gprop}) for accounting the Pauli blocking effect. The unfilled circle is the verticies coming from the in-vacuum subleading $\pi N$ chiral Lagrangian $\mathcal{L}_{\pi N}^{(2)}$.
It is notable that all the $O(p^{4})$ contributions to the correlation function $D^{ab}(p)$ vanishes in the soft limit. To see it,  
we evaluate the central diagram in Fig.~\ref{density} (a), as an example, with the external momentum $q_\mu$ using the in-medium propagator (\ref{Gprop}) and the vertices given in Appendix \ref{app}:
\begin{eqnarray*}
&& (-i)^2 \lim_{q \to 0} (2ifB_0 )^2 \Big{(} iD_\pi (q) \Big{)}^2 (-1)
\\ && \times
\int \frac{d^4 p}{(2 \pi )^4} {\rm Tr} \Big{[} (-i \frac{g_A}{2f} i \slash{q} \gamma_5 \tau^a )
 \\ && \times 
 iD^{-1}_m (p+q) (-i \frac{g_A}{2f} i \slash{q} \gamma_5 \tau^b ) iD^{-1}_m (p) \Big{]} \\
&=& 0 .
\end{eqnarray*}
Having taking the soft limit in the final equality, we find that the contribution of this diagram vanishes 
since the term in the trace becomes zero in the limit, while 
the pion propagator $D^{-1}(q)$ is finite thanks to the nonzero denominator. 
Since the other diagrams in Fig.~\ref{density} (a) have the same structure, 
we find that all the $ O(p^4) $ terms vanish in the soft limit and 
do not contribute to the  in-medium chiral condensate.
The reason that these terms vanish is that the pion is not a zero-mode 
off the chiral limit and the interaction between pion and nucleon is $p$-wave in the leading 
order.
We will see that $ \Pi^{ab}_5 (0)$ also have the same momentum dependence as the
leading term of $D^{ab}(0)$ up to  the NLO corrections in Sec.~\ref{cancellation}.
Thus, $ \Pi^{ab}_5 (0)$ does not contribute to the in-medium chiral condensate.

In Fig.~\ref{density} (b), we show all the diagrams of the leading order (LO) contribution 
in the density expansion, in which there are three diagrams.
We write the LO contribution for the in-medium chiral condensate as $\langle \bar{u}u + \bar{d}d \rangle^*_{\rm LO}$ and evaluate the diagrams in Fig.~\ref{density} (b) by the expansion of $1/m_N$ at the final state.
In the following we first calculate the left diagram in Fig.~\ref{density} (b) in the soft limit of the external momentum $q_\mu$:
\begin{eqnarray}
D^{ab}_{{\rm LO} 1} (0)
 &=& (-i)^2 \lim_{q \to 0}  \Big{(}  (2i f B_0) iD_\pi (q) \delta^{ab} \Big{)} 
 \nonumber \\ && \times (-1)
  \int \frac{d^4 p}{(2 \pi )^4} {\rm Tr} \Big{[} iD^{-1}_m (p) \frac{8i c_1 B_0}{f} \Big{]} \\
&=& -\frac{16 i c_1 B_0^2}{m_\pi^2} \delta^{ab}  \left(\Sigma_{p}^{1}(0) + \Sigma_{n}^{1}(0) \right)  \label{eq1} \\
&=& -\frac{16 i c_1 B_0^2 }{m_\pi^2} \delta^{ab} \rho \left( 1- \frac{3 k_F^2}{10 m_N^2} \right) .\label{eq2} 
\end{eqnarray}
Here  we have used the result of the tadpole nucleon loop $\Sigma_{N_{i}}^{1}(k)$ given in Eq.~\eqref{eq:tadpole}, in which we have expanded the result in terms of $1/m_{N}$ and taken the first two terms, and assumed the symmetric nuclear matter by taking $k_{F}^p = k_{F}^n =k_F$.

Next we calculate the central and right graphs in Fig.~\ref{density} (b). These contributions
denote $D^{ab}_{{\rm LO} 2} (0)$ and $D^{ab}_{{\rm LO} 3} (0)$, respectively: 
\begin{eqnarray}
\lefteqn{D^{ab}_{{\rm LO} 2} (0)} && \nonumber \\
 &=& (-i)^2 \lim_{q \to 0}  (2i f B_0)^2  \Big{(} iD_\pi (q) \Big{)}^2 \delta^{ab} 
 (-1) \int \frac{d^4 p}{(2 \pi )^4} {\rm Tr} \Big{[} 
 \nonumber \\&& 
 iD^{-1}_m (q) (-i) \Big{(} \frac{8B_0 c_1 m_q}{f^2} + \frac{4c_2}{f^2 m_N^2} (q \cdot p)^2 -\frac{2c_3}{f^2} q^2 \Big{)} \Big{]}
  \nonumber \\
&=& -\frac{32 i B_0^3 c_1 m_q}{m_\pi^4}  \delta^{ab} \int \frac{d^4 p}{(2 \pi )^4} {\rm Tr} \Big{[} iD^{-1}_m (q) \Big{]}  \\
&=& \frac{16 i B_0^2 c_1 }{m_\pi^2}  \delta^{ab} \rho \left( 1- \frac{3 k_F^2}{10 m_N^2} \right) .
\end{eqnarray}
Here we have used again the calculation of the tadpole nucleon loop~\eqref{eq:tadpole}
and the $k_{F}$ expansion has been made up to $k_{f}^{5}$. 
We have also used the relation $  m_\pi^2 = 2 m_q B_0 $. 
It is important to notice 
that the contributions coming from the $c_{2}$ and $c_{3}$ low energy constants
do vanish in the soft limit. $D^{ab}_{{\rm LO} 3} (0)$ can be calculated in the same way
as $D^{ab}_{{\rm LO} 1} (0)$
\begin{equation}
D^{ab}_{{\rm LO} 3} = D^{ab}_{{\rm LO} 1} = -\frac{16 i c_1 B_0^2 }{m_\pi^2} \delta^{ab} \rho \left( 1- \frac{3 k_F^2}{10 m_N^2} \right) .
\end{equation}

The leading contribution in the density expansion $D^{ab}_{\rm LO}$ is given by the sum of $D^{ab}_{\rm LO1} , D^{ab}_{\rm LO2}, D^{ab}_{\rm LO3}$.
We obtain the linear density contribution  of the in-medium condensate
together with the Fermi motion correction up to $k_{f}^{5}$:
\begin{eqnarray}
D^{ab}_{\rm LO} &=&  D^{ab}_{\rm LO1} + D^{ab}_{\rm LO2} + D^{ab}_{\rm LO3} \\
&=&  -\frac{16 i  B_0^2 c_1 }{m_\pi^2} \delta^{ab} \rho \left( 1- \frac{3 k_F^2}{10 m_N^2} \right) .
\end{eqnarray}
With this result we obtain the in-medium condensate in the normalization of 
the in-vacuum condensate as 
\begin{equation}
\frac{ \langle \bar{u}u + \bar{d}d \rangle^*_{\rm LO}}{\langle \bar{u}u + \bar{d}d \rangle_0} = \frac{4c_1}{f^2} \rho \left( 1- \frac{3 k_F^2}{10 m_N^2} \right) .
\label{LO}
\end{equation}
Here $c_1$ is one of the low energy constants (LECs) in $\mathcal{L}_{\pi N}^{(2)}$ 
and can be determined by  the $\pi N$ sigma term $\sigma_{\pi N}$.

  \begin{figure}[tb]
    \centering
    \includegraphics[width=5cm]{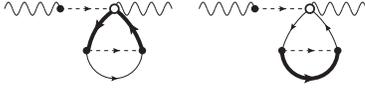}
    \caption{    Two examples of the diagrams  which have ultraviolet divergences and are counted as higher chiral orders but with the linear density order. }
    \label{p5correction}
  \end{figure}
  
As we mentioned before, we presume that the quantum corrections in vacuum is 
already accounted into the low energy constants and the renormalization 
procedure is supposed to be done. Thus, the loop contributions are renormalized 
into the low energy constants in the Lagrangian which we consider and 
we do not have to calculate the in-vacuum loop contributions in the in-medium calculation. 
For instance, we show two diagrams which have ultraviolet divergences and are 
to be counted as higher chiral orders while with the same density counting
in Fig.~\ref{p5correction}. The left diagram is the loop correction for the nucleon mass
and the right accounts the vertex correction. Both loop corrections are calculated
by the in-vacuum quantities. Therefore, we suppose that these loop corrections 
should be accounted in the nucleon mass and the vertex, respectively, and 
we use the observed values for these quantities. If one would follow the chiral counting 
scheme strictly, one would expand the physical quantities in terms of the chiral order
and discard higher order corrections than the order which one considers. 
Here we do not take the strict rule for the chiral counting. We exploit the
observed value, in which all order of the corrections should be included.
 
In this way, we fix the $c_{1}$ parameter by the observed $\pi N$ sigma term as
\begin{equation}
\sigma_{\pi N} = - 4 c_1 m_{\pi}^2 ,
\label{sigma}
\end{equation}
admitting that the loop corrections for the $\pi N$ sigma term are taken 
into account into the physical value, and we do not calculate further 
the in-vacuum loop contribution, most of which are divergent.  
Therefore, the leading order contribution of the in-medium chiral condensate 
in the density expansion is given by the in-vacuum physical quantities as
\begin{equation}
\frac{ \langle \bar{u}u + \bar{d}d \rangle^*_{\rm LO}}{\langle \bar{u}u + \bar{d}d \rangle_0} =  - \frac{\sigma_{\pi N}}{f_\pi^2 m_\pi^2} \rho \left( 1- \frac{3 k_F^2}{10 m_N^2} \right).
\label{LO}
\end{equation}
The second term is  a $ 1/m_N $ correction to the linear density and it is counted as NNLO $ O(p^7) \sim O(\rho^{5/3}) $. To complete the $O(\rho^{5/3})$ contributions, we have 
to calculate further higher orders in the density expansion, as we shall see later. 
Here it is important to emphasize again that we take the observed values. 
The values which we use in this work are the sigma term 
$\sigma_{\pi N} = 45 {\rm MeV} $ \cite{GLS}, the pion decay constant $f_\pi=92.4 {\rm MeV}$, the pion mass $ m_\pi = 138 {\rm MeV}$ and the nucleon mass $m_N = 938 {\rm MeV}$.
The result \eqref{LO} coincides the well-known linear density approximation result \cite{Coh91,DL}.

Now we evaluate the NLO density corrections $ \langle \bar{u}u + \bar{d}d \rangle^*_{\rm NLO} $. 
The relevant diagrams for the next leading order are shown in Fig.~\ref{density} (c).
These diagrams contain two loops coming from the nucleon in the Fermi sea and 
a free pion. The diagrams in which one of the nucleon propagators is 
the free propagator is already accounted as higher chiral order terms
into the renormalized $\pi N$ vertex, because the diagram contains a loop 
written by only the free propagators, which is divergent and should be renormalized
into the in-vacuum vertex. 
Here we use such a parameterization 
of the chiral field $U$ in terms of the pion field that the naive perturbative
expansion can be done. The details are written in Appendix \ref{para}. 
The leftmost diagram in third row in Fig.~\ref{density} is calculated as
\begin{eqnarray*}
&& \frac{1}{2} (-i)^2 \lim_{q \to 0} \sum_{i j} (-1) \int \frac{d^4 k}{(2 \pi )^4} \frac{d^4 p}{(2 \pi )^4} (2i f B_0) iD_\pi (q) 
\\ && \times 
\left( -\frac{2iB_0}{5f} \big{(} \delta^{ai} \delta^{bj} +\delta^{aj} \delta^{ib} +\delta^{ab} \delta^{ij} \big{)}  \right) \left( iD_\pi (k) \right)^2 \\
&& \times {\rm Tr} \Big{[} ( \frac{ig_A}{2f} i\slash{k} \gamma_5 \tau^i ) iD^{-1}_m (p+k) (-\frac{ig_A}{2f} i\slash{k} \gamma_5 \tau^j )  iD^{-1}_m (p) \Big{]}  \\
&=&\delta^{ab}\int \frac{d^4 k}{(2 \pi )^4}  \frac{-i g_A^2 B_0^2}{f^2 m_{\pi}^2}   \left( \frac{1}{k^2- m_{\pi}^2 +i \epsilon } \right)^2 \Sigma_{N}^{2}(k) \\
&=& -\frac{2 i g_A^2 B_0^2}{ f^2 m_{\pi}^2} \delta^{ab} \frac{k_F^4}{6 \pi^4} F(\frac{m_\pi^2}{4k_F^2}) .
\end{eqnarray*}
where the factor $1/2$ comes from the symmetric factor for the loop. The calculation of the nucleon one-loop in the Fermi sea is given 
in Eq.~\eqref{eq:fermiloop}. We have defined
\begin{eqnarray}
F(a^2)&=&\int_{0}^{1}dx \left(\frac{x^{2}}{x^{2}+a^2}\right)^{2} \frac{1}{2} (1-x)^{2} (x+2)\\
&=&  \frac{3}{8} -\frac{3a^2}{4} -\frac{3a}{2} {\rm arctan} \frac{1}{a}
 + \frac{3a^2}{4} (a^2 +2) \ln \frac{a^2 +1}{a^2} . \nonumber 
\end{eqnarray}
After we evaluate the other diagrams in the third row in Fig.~\ref{density}, we obtain
\begin{eqnarray}
\frac{ \langle \bar{u}u + \bar{d}d \rangle^*_{\rm NLO 1}}{\langle \bar{u}u + \bar{d}d \rangle_0} 
= \frac{g_A^2 k_F^4}{4 f_\pi^4 \pi^4} F(\frac{m_\pi^2}{4k_F^2})  .
\label{NLO1}
\end{eqnarray}
Here, we also take the physical value of the axial coupling $g_A = 1.27$.
This term is a density correction to the $\pi N$ sigma term through the pion-loop and is proportional to $ \rho^{4/3} $.

The Feynman diagrams in the fourth row in Fig.~\ref{density} also contribute to 
the next-to-leading order,
representing the Ericson-Ericson double-scattering correction \cite{Eri66}.
For example, the middle diagram of the fourth two in Fig.~\ref{density}
can be calculated as
\begin{eqnarray}
&-& \lim_{q \to 0} (2if B_0)^2 ( iD_\pi (q) )^2 \left( \frac{4B_{0}c_{1} m_{q}}{f^{2}}\right)^{2} \nonumber \\ 
&& \times \int \frac{d^{4}k}{(2\pi)^{4}} \frac{d^{4}p}{(2\pi)^{4}} {\rm Tr} \Big{[} iD_{m}^{-1}(q-\frac{p}{2}) \nonumber \\
&& \times iD_{m}^{-1}(q+\frac{p}{2}) \Big{]} i D_{\pi}(p+k)  \label{doublescat} \\
  &=& \frac{2 B_0^2 \sigma_{\pi N}^2 k_F^4}{3 f_\pi^2 \pi^4 m_\pi^4 } \Sigma_{N}^{4}(0) \\
  &=& \frac{-8i B_0^2 \sigma_{\pi N}^2 k_F^4}{3 f_\pi^2 \pi^4 m_\pi^4 } G(\frac{m_\pi^2}{4k_F^2}) ,
\end{eqnarray}
where $G(a)$ is given as 
\begin{displaymath} 
G(a^2) =  \frac{3}{8} - \frac{a^2}{4} -a \arctan \frac{1}{a} + \frac{a^2}{4} (a^2 +3 ) \ln |\frac{1+a^2}{a^2}| .
\end{displaymath}
The detail calculations are summarized in Appendix \ref{loops}.
In Eq.~\eqref{doublescat} we have understood that the terms with 
$c_{2}$ and $c_{3}$ vanish at the soft limit because they are proportional to 
the pion momentum $q_{\mu}$. 
Including the rest of the diagram for the double scattering correction, we obtain
\begin{eqnarray}
\frac{ \langle \bar{u}u + \bar{d}d \rangle^*_{\rm NLO 2}}{\langle \bar{u}u + \bar{d}d \rangle_0}
&=& \frac{2 \sigma_{\pi N}^2  k_F^4}{3 f_\pi^4 \pi^4 m_\pi^2 } G(\frac{m_\pi^2}{4k_F^2})  .
\label{NLO2}
\end{eqnarray}

In this way, we obtain NLO contributions of in-medium chiral condensate:
\begin{eqnarray}
&& \frac{ \langle \bar{u}u + \bar{d}d \rangle^*_{\rm NLO}}{\langle \bar{u}u + \bar{d}d \rangle_0}
=  \frac{ \langle \bar{u}u + \bar{d}d \rangle^*_{\rm NLO 1}}{\langle \bar{u}u + \bar{d}d \rangle_0}  + \frac{ \langle \bar{u}u + \bar{d}d \rangle^*_{\rm NLO 2}}{\langle \bar{u}u + \bar{d}d \rangle_0}  \nonumber \\ 
&=& \frac{g_A^2 k_F^4}{4 f_\pi^4 \pi^4} F(\frac{m_\pi^2}{4k_F^2}) + \frac{2 \sigma_{\pi N}^2  k_F^4}{3 f_\pi^4 \pi^4 m_\pi^2 } G(\frac{m_\pi^2}{4k_F^2})  .
\label{NLO} 
\end{eqnarray}

From Eqs.~(\ref{LO}) and (\ref{NLO}), we obtain the in-medium chiral condensate within NNLO corrections($O (\rho^{5/3} )$)
\begin{eqnarray}
\frac{ \langle \bar{u}u + \bar{d}d \rangle^*}{\langle \bar{u}u + \bar{d}d \rangle_0} &=& 1 - \frac{\sigma_{\pi N}}{f_\pi^2 m_\pi^2} \rho \left( 1- \frac{3 k_F^2}{10 m_N^2} \right) \nonumber \\
& +& \frac{g_A^2 k_F^4}{4 f_\pi^4 \pi^4} F(\frac{m_\pi^2}{4k_F^2}) + \frac{2 \sigma_{\pi N}^2  k_F^4}{3 f_\pi^4 \pi^4 m_\pi^2 } G(\frac{m_\pi^2}{4k_F^2}). \nonumber \\
\label{chiconNLO}
\end{eqnarray}

Let us comment on the $\Delta$ resonance contribution up to $O(\rho^{5/3})$. 
The $\Delta$ resonance contributes to the in-medium amplitudes through the 
$\Delta$-hole excitation in these orders. Nevertheless, the $\Delta$-hole 
excitation in the left diagram of Fig.~\ref{density} (a) vanishes in the soft limit
because of the $p$-wave nature of the $\pi N \Delta$ coupling 
as we have seen in the nucleon-hole excitation. The $\Delta$-hole excitation 
in the diagrams of Fig.~\ref{density} (c) should be accounted in the in-vacuum 
$P\pi NN$ vertex, since the loop contribution of the pion and $\Delta$ in this 
diagram appears in the higher order calculation of the $P\pi NN$ vertex 
in the chiral expansion. Therefore, up to NLO there are no explicit $\Delta$ 
contributions to the in-medium chiral condensate.

\subsection{Cancellation of $ \Pi^{ab}_5 (0)$ }
\label{cancellation}

 \begin{figure}[t]
    \centering
    \includegraphics[width=8cm]{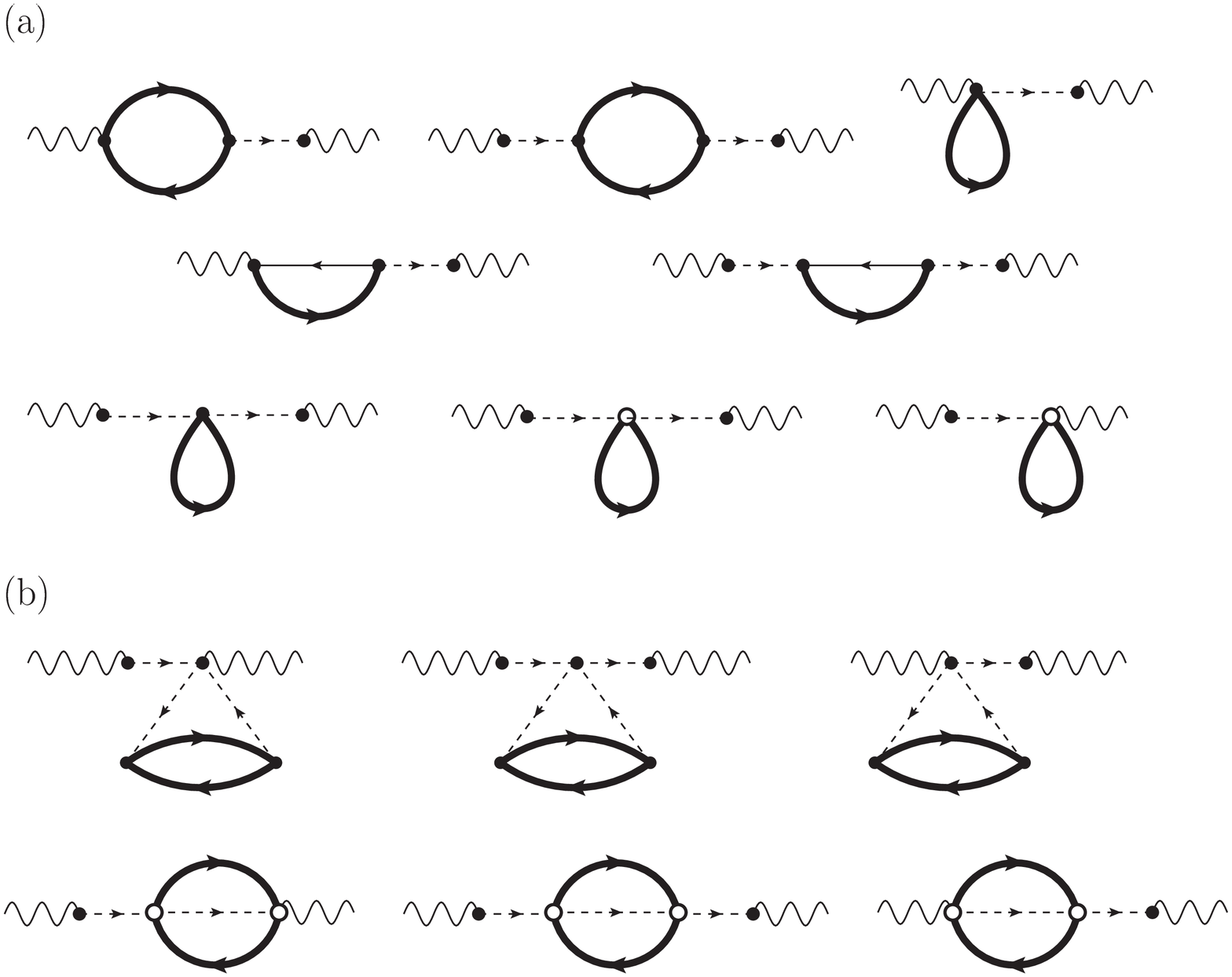}
    \caption{Feynman diagrams for the $ \Pi^{ab}_5 (q)$ correlation function:
    the leading density contribution (a) 
    and  the next to leading density corrections (b).
    The left and right wavy lines 
express the axial and pseudoscalar currents, respectively, 
the dashed line denotes the pion propagation and the solid dots stand  for the leading vertices given by the chiral Lagrangian $\mathcal{L}_{\pi}^{(2)}$.
}
    \label{vanish}
  \end{figure}

According to Eq.~\eqref{ward}, $ \Pi^{ab}_5 (0)$ can also contribute to the
in-medium condensate. Nevertheless, the correlation function $ \Pi^{ab}_5 (q)$
gives a null value in the soft limit off the chiral limit. In this section,
we confirm the cancellation of $ \Pi^{ab}_5 (q)$ in the soft limit. 

In vacuum the correlation function $ \Pi^{ab}_5 (q)$ should vanish
in the soft limit because there is no zero mode propagation
off the chiral limit and the coupling of the axial current to pseudoscalar 
modes is proportional to the external momentum $q$, which is taken
to be zero in the soft limit. This can be seen  in the leading order 
calculation of the in-vacuum contribution as
\begin{eqnarray}
\Pi_{50}^{ab}(q) &=&
q^\mu i (iq_\mu f) \frac{i}{q^2 - m_\pi^2 + i\epsilon} i(2fB_0) \nonumber \\ 
&  \to &  0 \ \ {\rm for} \ \ q \to 0. 
\end{eqnarray}

For the in-medium contributions of the correlation function $ \Pi^{ab}_5 (q)$
we show the Feynman diagram $ \Pi^{ab}_5 (p)$ in Fig.~\ref{vanish}. 
In this figure, the left and right wavy lines express the axial and 
pseudoscalar currents, respectively, and
the dashed line denotes the pion propagation. 
In the linear density order, we also find that $ \Pi^{ab}_5 (q)$ vanishes in the soft limit.
For example, we evaluate the upper left diagram in (a):
\begin{eqnarray*}
\lefteqn{ \Pi_{5 {\rm LO1}}^{ab}(0)} \nonumber && \\
& = & \lim_{q \to 0} q^\mu iD_\pi (q) (2ifB_0) (-1)
\int \frac{d^4 p}{(2 \pi )^4} {\rm Tr} \Big{[} (i g_A i \gamma_\mu \gamma_5 \frac{\tau^a}{2} )
   \nonumber \\ && \ \ \ \ \ \ \  \times 
 iD_{m}^{-1}(p+q) (-i \frac{g_A}{2f} i \slash{q} \gamma_5 \tau^b ) i D_{m}^{-1}(p) \Big{]} \\
&=& \lim_{q \to 0} D_\pi (q) (- i) g_{A}^{2} B_{0} \delta^{ab} \Sigma^{2}_{N}(q)  = 0 .
\end{eqnarray*}
where we have used the one-nucleon loop function given in Eq.~\eqref{eq:fermiloop},
and in the last equation we have used the fact that the pion propagator is finite 
in the soft limit owing to the finite pion mass and the nucleon one-loop vanishes.
This cancellation comes from two reasons;
the one is that pion is not a zero mode and the other is that interaction between pion and the axial current is derivative interaction, in other words
explicit and spontaneous symmetry breaking leads to this cancellation.
In addition, the interactions between pion and nucleon is proportional to the pion momentum because of spontaneous symmetry breaking, and it causes the same result.
In the same manner one can confirm that contributions coming from the other diagrams give null value 
in the soft limit.

Up to the NLO density order, we find that the $ \Pi^{ab}_5 (0)$ correlation function vanishes.
Generally speaking, there exist zero modes which couple with the axial current such as one-particle one-hole excitation as discussed in \cite{JHK}. Nevertheless, up to the NLO corrections we find that such zero modes do not contribute.

We note that in the chiral limit the $ \Pi^{ab}_5 (0)$ correlation function contributes the in-medium chiral condensate, while the $D^{ab}(0) $ correlation function vanishes.  One can find easily that the momentum dependence of the pion propagator cancels to the external momentum in the soft limit
and $ \Pi^{ab}_5 (0)$ remains finite. The in-medium chiral condensate in the soft limit
calculated by the $ \Pi^{ab}_5$ correlation function reads
\begin{equation}
\frac{ \langle \bar{u}u + \bar{d}d \rangle^*_{LO}}{\langle \bar{u}u + \bar{d}d \rangle_0} = \frac{4 c_1}{ f_\pi^2} \rho \left( 1- \frac{3 k_F^2}{10 m_N^2} \right) 
\end{equation}
for the leading order of the density expansion and
\begin{eqnarray*}
\frac{ \langle \bar{u}u + \bar{d}d \rangle^*_{NLO}}{\langle \bar{u}u + \bar{d}d \rangle_0} =  \frac{3 g_A^2}{32 \pi^2 f^4} \left( \frac{3\pi^2}{2} \right)^{\frac{1}{3}} \rho^{\frac{4}{3}}
\end{eqnarray*}
for the next leading order. These are equivalent to the result obtained from $D^{ab}(0) $ 
off the chiral limit by taking the chiral limit afterwards.

\subsection{Density dependence of chiral condensate}
\label{densitydependence}

\begin{figure}[t]
    \centering
    \includegraphics[width=6.0cm,angle=-90]{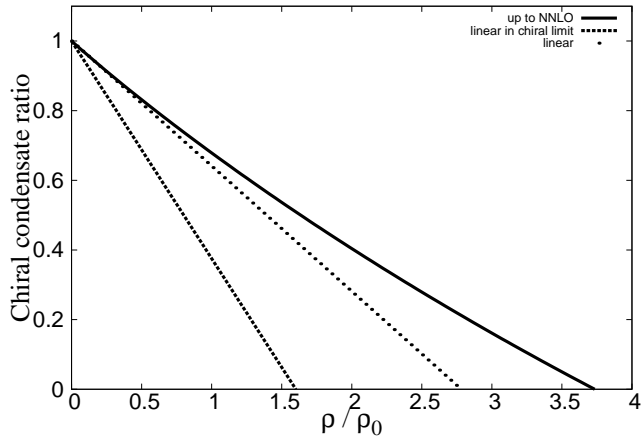}
    \caption{The density dependence of the ratio of the chiral condensates,
   $\langle \bar{u}u + \bar{d}d \rangle^* / \langle \bar{u}u + \bar{d}d \rangle_0 $, as 
   a function of $\rho / \rho_0$ in symmetric nuclear matter. The dashed, dotted and solid lines represent the linear density result in the chiral limit,  the linear density result off chiral limit and the  result up to the next leading order of the density expansion off the chiral limit, respectively.}
    \label{chiral2}
  \end{figure}

In Fig.~\ref{chiral2}, the density dependence of the ratio of the chiral condensates,
$\langle \bar{u}u + \bar{d}d \rangle^* / \langle \bar{u}u + \bar{d}d \rangle_0 $, is plotted as a function of $\rho / \rho_0$ in symmetric nuclear matter.
The solid line represents the next-leading order (NLO) result shown in Eq.~\eqref{chiconNLO}.
For comparison, we also show the linear density results in and off the chiral limit
as the dashed and dotted lines, respectively. For the result in the chiral limit, we have 
used chiral limit values of the $c_{1}$ parameter $c_1 \approx 0.93 {\rm GeV^{-1}} $ and the pion decay constant $f \approx 88 {\rm MeV} $.

We find that the linear density result in the chiral limit decreases more rapidly than the results off chiral limit. The NLO correction amounts to about as small as 5\% at $\rho = \rho_0$, and becomes
significant in higher density, for instance, at $\rho = 2\rho_0$ the NLO correction is around 
10 \%. 
Therefore the linear density approximation is good in low densities, while in higher density 
the NLO contribution is not ignorable.
Numerically, we find that at normal nuclear density 
up to LO $\langle \bar{u}u + \bar{d}d \rangle^* / \langle \bar{u}u + \bar{d}d \rangle_0 \approx 0.65$
and up to NLO $\langle \bar{u}u + \bar{d}d \rangle^* / \langle \bar{u}u + \bar{d}d \rangle_0 \approx 0.68$.
These values are close to the value suggested by the recent precise pionic atom determination $\langle \bar{u}u + \bar{d}d \rangle^* / \langle \bar{u}u + \bar{d}d \rangle_0 \approx 0.67$ \cite{Su}.
We note that this experimental value is determined by linear density extrapolation 
under the assumption that the pion bound in the $1s$ orbit is in a nuclear medium with 
an effective density $\rho_{e} \approx 0.6 \rho_{0}$.  
We also evaluate the quark condensate at $\rho = 0.6 \rho_0$ 
and find  $\langle \bar{u}u + \bar{d}d \rangle^* / \langle \bar{u}u + \bar{d}d \rangle_0 \approx 0.78$ for LO
and $\langle \bar{u}u + \bar{d}d \rangle^* / \langle \bar{u}u + \bar{d}d \rangle_0 \approx 0.80$
up to NLO. 
These values are very close to the experimentally extracted value of the ratio $b_{1}^{\rm free}/b_{1}=0.78\pm 0.05$~\cite{Su}, where $b_{1}$ is a parameter of the optical potential for the in-medium pion presenting the in-medium isovector $\pi N$ scattering length and $b_{1}^{\rm free}$ is the $\pi N$ isovector scattering length. 
Under the linear density approximation and a small isoscalar $\pi N$ scattering length, 
the ratio of  $b_{1}^{\rm free }/b_{1}$ is equivalent to the ratio of the chiral condensate. 
Thus, this implies that the density expansion might be good in at least low density region.

We note that in-medium CHPT is a low energy effective theory and this theory would be applicable up to
about 2 normal density because at twice the normal nuclear density Fermi momentum is about $340 {\rm MeV}$.
In further higher density region this theory will be beyond applicability.
Nevertheless we could estimate the density at which chiral symmetry is restored.
In the Fig.~\ref{chiral2}, we would find that the NLO correction raises 
the symmetry restoration density from $3\rho_0$ to $4\rho_0$, which 
would imply that the NLO correction is not ignorable in high density region.

\subsection{Higher order corrections and role of $NN$ contact terms}
\label{contactNN}

  \begin{figure}[h]
    \centering
    \includegraphics[width=6cm]{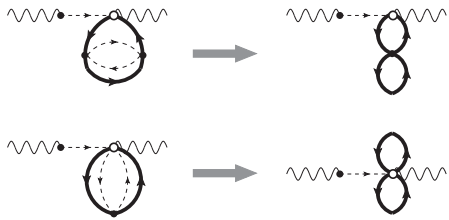}
    \caption{Contact terms for renormalization of higher order corrections beyond NLO.}
    \label{contact}
  \end{figure}

When one considers further higher order correction beyond NLO, one encounters 
divergent amplitudes even though all of the $\pi N$ interactions is fixed in vacuum. 
For instance, diagrams of multi-pion exchange in Fermi gan as shown in Fig.~\ref{contact}
give divergent amplitudes. In the figure the solid line denotes Fermi see insertion and 
the dashed line represents pion propagation. These diagrams count as $O(k_{f}^{6})$
namely $O(\rho^{2})$. The upper diagram is proportional to
\begin{eqnarray}
&&  \int \frac{d^4 p'}{(2 \pi )^4} \frac{d^4 p}{(2 \pi )^4} \frac{d^4 k}{(2 \pi )^4} {\rm Tr} \Big{[} (-i A_{\pi P}^{(1)} ) iD_m^{-1} (p)  (-i A_{\pi \pi}^{(1)} ) 
\nonumber \\ && \times
iD_m^{-1} (k)  (-i A_{\pi \pi}^{(1)} ) iD_m^{-1} (p)\Big{]} 
iD_\pi (p') iD_\pi (p+p' -k) . \nonumber 
\end{eqnarray}
In this expression the integral with respect to $p^\prime$ for the pion loop gives divergence.
As pointed out in Ref.~\cite{O}, we need $NN$ contact terms to 
control the divergence.  
This means that one can proceed the in-medium calculation up to NLO by using the $\pi N$ dynamics, but if one considers $O(\rho^{2})$ and higher corrections, one needs  also in-vacuum $NN$ contact terms obtained by the $NN$ dynamics~\cite{Dmitrasinovic:1999pu}.
We emphasize that in order to evaluate the in-medium chiral condensate quantitatively
with higher density corrections, we need not only the $\pi N$ dynamics informations but also $NN$ dynamics. 
Recently, as a step in this direction, a non-perturbative chiral effective theory has been developed to improve the $NN$ correlation by including $NN$ contact terms using a resummation method \cite{LOM}.
Moreover, in Ref. \cite{KHW} the $\Delta$(1232) resonance contributions have been evaluated and it has been found that $\Delta$ resonance effects, which appears from the $O(\rho^{2})$ contributions,  are not small. 
Therefore, we may need more sophisticated calculations for the in-medium chiral condensate including $NN$ and more dynamics.


\section{Summary}
We calculate the chiral condensate at finite nuclear density $ \langle \bar{u}u + \bar{d}d \rangle^* $ using the chiral Ward identity and the in-medium chiral perturbation theory.
We study diagrammatic structure of the current-current Green's functions $\langle \Omega | T A_\mu^a(x) P^b(0) | \Omega \rangle$ and $\langle \Omega | T P^a_{\mu}(x) P^b(0) | \Omega \rangle$ and classify density corrections to the chiral condensate.
In our study we fix the $\pi N$ dynamics in vacuum and calculate the in-medium 
chiral condensate with the in-medium chiral perturbation theory.
In this study, LO($O(\rho)$) reproduces the well-known linear density approximation to 
the chiral condensate.
This leading density correction is proportional to the $\pi N $ sigma term $\sigma_{\pi N} $. The next leading correction, NLO($O(\rho^{4/3})$),
represents in-medium corrections of the sigma term.
As a results, we find that linear density approximation is rather good in low density region such as a normal nuclear density.
We have found that for higher corrections the correlation function has divergence 
from the pion loop even though all the in-vacuum quantities for the $\pi N$ dynamics 
are fixed. 
This means that
the $O(\rho^{2})$ corrections  cannot be determined only by the $\pi N$ dynamics in 
vacuum and the information of the $NN$ dynamics is necessary to control the divergence. 
It should be also emphasized that for a realistic nuclear matter one should incorporate the NN dynamics and check that the matter satisfies the nuclear matter properties.


\section*{Acknowledgement}
This work was partially supported by the Grants-in-Aid for Scientific Research (No.\ 25400254 and No.\ 24540274).


\appendix
\section{Parametrization of the chiral field}
\label{para}
The chiral perturbation theory successfully introduces the chiral invariant 
Lagrangian in the spontaneous breaking of chiral symmetry. The chiral field $U$ 
transforming linearly under the chiral rotation is written nonlinearly in terms of
the pion field. The parametrization of the chiral field in terms of the pion field
is not unique~\cite{Weinberg:1968de} and all of the correct parameterizations 
provide the same physical result. 
However, one should realized that the basic field for the nonlinear sigma model 
to maintain chiral symmetry is the chiral field $U$ not the pion field as 
one can see that the partition function of the chiral perturbation theory is defined 
by the path integral with respect to the chiral field $U$. Therefore, when 
one considers quantum corrections of the pion field in perturbative calculations, 
one should be careful with chiral invariance, naive perturbative calculations 
might break chiral symmetry~\cite{Charap:1971bn}. 
The perturbative expansion of the pion fields is defined 
by the path integral with respect to the pion field, so that one should make 
the integral measure to be chiral invariant~\cite{Gerstein:1971fm}. 
One of the popular prescription is the background field method
as it was applied to the chiral perturbation theory in Ref.~\cite{GL1}.

Here, instead of the cerebrated CCWZ parametrization,  
we take the parametrization of the $U$ field which can be used 
for the naive perturbative calculation. This was found in 
Refs.~\cite{Charap:1971bn,Gerstein:1971fm}.  In this parametrization 
the chiral field is written \cite{Gerstein:1971fm} as
\begin{equation}
   U = \exp\left[ i \pi^i \tau^i \frac{y(\pi^{2})}{2 \sqrt{\pi^{2}}} \right]
\end{equation}
where the function $y(\pi^{2})$ satisfies
\begin{equation}
   y - \sin y = \frac{4}{3} \left(\frac{\pi^{2}}{f^{2}} \right)^{\frac{3}{2}}
\end{equation}
For the calculation we expand the chiral field in terms of the pion field \cite{Charap:1971bn} as
\begin{equation}
U= 1+ \frac{i \pi^i \tau^i}{f} - \frac{\pi^i \pi^i }{2 f^2} - \frac{i \pi^i \tau^i \pi^j \pi^j }{10 f^3} - \frac{\pi^i \pi^i \pi^j \pi^j }{40 f^4} + \cdots \label{expansion}
\end{equation}
and take some first terms.

\section{Chiral Lagrangian and $\pi$N interaction}
\label{app}
In this section, we show the chiral Lagrangian and the $\pi N$ interaction 
which we use in this work. The chiral Lagrangian for the pion sector is as follows:
\begin{equation}
\mathcal{L}_{\pi}^{(2)} = \frac{f^2}{4} {\rm Tr} \left( D_\mu U^\dag D^\mu U + \chi^\dag U + \chi U^\dag \right)  \label{Lag}
\end{equation}
where the covariant derivative is defined with the vector external fields as
\begin{equation}
D_\mu U \equiv \partial_{\mu} U -i ( v_{\mu} + a_{\mu} )U +iU( v_{\mu} - a_{\mu} ) 
\end{equation}
and the external scalar fields are given by
\begin{equation}
\chi = 2B_0 (s + ip) .
\end{equation}
In the following we list up the interaction Lagrangian 
relevant for the present calculations. These terms are obtained from the Lagrangian 
\eqref{Lag} with the expansion~\eqref{expansion}:
\begin{itemize}

\item $\pi$-$P$ vertex:
\begin{equation}
\mathcal{L}^{(2)}_{\pi P} = 2f B_0 \pi^i P^i
\end{equation}

\item $\pi \pi \pi $-$P$ vertex:
\begin{equation}
\mathcal{L}_{\pi^3 P}^{(2)} = -\frac{B_0}{5f} P^i  \pi^i \pi^j \pi^j 
\end{equation}

\item $\pi \pi \pi \pi$ vertex:
\begin{eqnarray}
\mathcal{L}^{(2)}_{\pi^4} &=& - \frac{1}{10 f^2} \partial_\mu \pi^i \partial^\mu \pi^j \pi^k \pi^l (\delta^{i j} \delta^{kl} \nonumber  \\
&& - 3 \delta^{ik} \delta^{jl}  ) - \frac{m_q B_0}{20 f^2} \pi^i \pi^j \pi^k \pi^l \delta^{ij} \delta^{kl} 
\end{eqnarray}

\item $\pi \pi \pi $-$a_{\mu}$ vertex:
\begin{equation}
\mathcal{L}^{(2)}_{\pi^3 a} = \frac{1}{5f} a_\mu^i \partial^\mu \pi^j \pi^k \pi^l (3\delta^{ij} \delta^{kl} - 4 \delta^{ik} \delta^{jl})
\end{equation}
\end{itemize}

The chiral Lagrangian for the one-nucleon sector is as follows:
\begin{equation}
\mathcal{L}_{\pi N} =  \bar{N} (i \gamma^\mu \partial_\mu -m_N -A) N ,
\end{equation}
where $A$ represents all the chiral interaction with the nucleon bilinear form 
and can be counted in terms of the pion momentum:
\begin{displaymath}
A = \sum_{n=1} A^{(n)} 
\end{displaymath}
Here $ A^{(n)} $ is counted as $O(p^n)$. 

The explicit form of  the leading term $ A^{(1)} $ is 
\begin{equation}
A^{(1)} = -i \gamma^{\mu} \Gamma_{\mu} - i g_A \gamma^{\mu} \gamma_5 \Delta_{\mu} 
\end{equation}
with the vector current
\begin{equation}
\Gamma_{\mu} = \frac{1}{2} [u^\dag , \partial_{\mu} u ] - \frac{i}{2} u^\dag ( v_{\mu} + a_{\mu} ) u - \frac{i}{2} u ( v_{\mu} - a_{\mu} ) u^\dag  
\end{equation}
and the axial current
\begin{equation}
\Delta_{\mu} = \frac{1}{2} \Big{[} u^\dag \big{(} \partial_\mu - i (v_\mu + a_\mu ) \big{)} u  - u \big{(} \partial_\mu - i(v_\mu - a_\mu ) \big{)} u^\dag \Big{]}
\end{equation}
Here we define $ u = \sqrt{U} $. 
The explicit expression of the next leading term $ A^{(2)} $ is 
is given as
\begin{eqnarray}
A^{(2)} &=& -c_1 \langle \chi_+ \rangle + \frac{c_2}{2m_N^2} \langle u_{\mu} u_{\nu} \rangle D^{\mu} D^{\nu} -\frac{c_3}{2} \langle u_{\mu} u^{\mu} \rangle \nonumber  \\
&& + \frac{c_4}{2} \gamma^{\mu} \gamma^{\nu} [u_{\mu}, u_{\nu} ] -c_5 \hat{\chi}_+ -\frac{ic_6}{8m_N} \gamma^{\mu} \gamma^{\nu} F_{\mu \nu}^+\nonumber \\
&& - \frac{ic_7}{8m_N} \gamma^{\mu} \gamma^{\nu} \langle F_{\mu \nu}^+ \rangle 
\end{eqnarray}
with 
\begin{eqnarray*}
&& D_\mu \psi = \partial_\mu \psi + \Gamma_\mu \psi \\
&& u_\mu = 2i \Delta_\mu \\
&& \chi_+ = u \chi^\dag u + u^\dag \chi u^\dag \\
&& \hat{\chi}_+ = \chi_+ - \frac{1}{2} \langle \chi_+ \rangle, \\
&& F_{\mu \nu}^+ = u^\dag F_{\mu \nu}^R u + u F_{\mu \nu}^L u^\dag \\
&& F_{\mu \nu}^R =\partial_\mu r_\nu - \partial_\nu r_\mu -i [r_\mu , r_\nu ] ,\hspace{2em} r_\mu = v_\mu + a_\mu \\
&& F_{\mu \nu}^L =\partial_\mu l_\nu - \partial_\nu l_\mu -i [l_\mu , l_\nu ] ,\hspace{2em} l_\mu = v_\mu - a_\mu
\end{eqnarray*}

We list up the vertices which we use in the calculation:
\begin{itemize}
\item  $aNN$ vertex: 
\begin{equation}
A_{a}^{(1)} = - g_A \gamma^\mu \gamma_5 a_\mu^i \frac{\tau^i}{2} 
\end{equation}
\item
$\pi aNN$ vertex:
\begin{eqnarray}
A_{\pi a}^{(1)} &=& \frac{i}{2f} \gamma^\mu [\pi, a_\mu ] = - \frac{1}{2f} \gamma^\mu \pi^i a_\mu^j \epsilon^{ijk} \tau^k \\
A_{\pi a}^{(2)} &=& -\frac{2c_2}{f m_N^2} \partial_{\mu} \pi^i a_{\nu}^i \partial^{\mu} \partial^{\nu} +\frac{2c_3}{f} \partial_{\mu} \pi^i a^{\mu i} \nonumber \\
&& - \frac{ic_4}{f} \epsilon^{ijk} \tau^k \partial_\mu \pi^i a_{\nu}^j [\gamma^\mu, \gamma^\nu ] \nonumber \\
\end{eqnarray}
\item
$\pi PNN$ vertex:
\begin{equation}
A_{\pi P}^{(2)} = -\frac{8 c_1 B_0}{f} P^i \pi^i
\end{equation}
\item
$\pi NN$ vertex:
\begin{equation}
A_{\pi}^{(1)} = \frac{g_A}{2f} \gamma^{\mu} \gamma_5 \partial_{\mu} \pi^i \tau^i 
\end{equation}
\item
$\pi \pi NN$ vertex:
\begin{eqnarray}
A_{\pi \pi}^{(1)} &=&  -\frac{i}{8f^2} \gamma^{\mu} [\pi ,\partial_{\mu} \pi ] = \frac{\gamma^{\mu}}{4f^2} \pi^i \partial_{\mu} \pi^j \epsilon^{ijk} \tau^k \\
A_{\pi \pi}^{(2)} &=& \frac{4B_0 c_1 m_q}{f^2}\pi^i \pi^i + \frac{c_2}{f^2 m_N^2} \partial_{\mu} \pi^i \partial_{\nu} \pi^i \partial^{\mu} \partial^{\nu}  \nonumber \\
&& -\frac{c_3}{f^2} \partial_{\mu} \pi^i \partial^{\mu} \pi^i +i \frac{ic_4}{f^2} \epsilon^{ijk} \tau^k \gamma^{\mu} \gamma^{\nu} \partial_{\mu} \pi^i \partial_{\nu} \nonumber 
\end{eqnarray}
\end{itemize}

\section{In-medium nucleon loops}
\label{loops}
In this section we calculate the nucleon loop diagrams in the Fermi sea. 
In these calculations we take the trace only in the spin space, which is 
indicated by ${\rm Tr}_{s}$, and the isospin will be
considered in other places.

First of all, we calculate the nucleon tadpole $ \Sigma^1_{N_i} (k)$ which appears in a diagram (a) shown in Fig.~\ref{sigma}:
\begin{eqnarray}
\Sigma^1_{N_i} (k) &=& -\int \frac{d^4 p}{(2 \pi )^4} {\rm Tr_s} \Big{[} iD^{-1}_m (p) \Big{]} \nonumber \\
&=&  \int \frac{d^3 p}{(2 \pi )^3 2 E({\bf p})} {\rm Tr_s} \Big{[} (\slash{p} + m_N)  \Big{]} \theta (k_F^i - |{\bf p}|) \nonumber \\
&=&  \int^{k_F^{(i)}}_{0} \frac{{\bf p}^2 dp}{\pi^2} \frac{m_{N}}{E({\bf p})}  \nonumber \\
&\approx &  \int^{k_{F}}_{0} \frac{{\bf p}^2 dp}{\pi^2} \left( 1- \frac{{\bf p}^2}{2 m_N^2} \right) \label{eq:nonrela} \nonumber \\
&=& \rho^i \left( 1- \frac{3 k_F^{i2}}{10 m_N^2} \right), \label{eq:tadpole}
\end{eqnarray} 
In Eq.~\eqref{eq:nonrela}, we have taken the first two terms in the $1/m_{N}$ expansion 
and the nucleon density is given by $\rho^i =  2 k_{F}^{i3} / (3\pi^{2})$. 

  \begin{figure}[t]
    \centering
    \includegraphics[width=8cm]{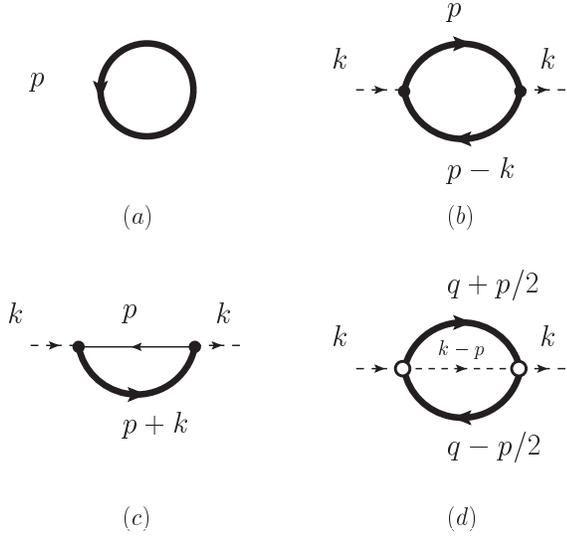}
    \caption{Nucleon loop diagrams. (a) the tadpole diagram,  $\Sigma_{N_{i}}^{1}(k)$. (b) the one-loop diagram of nucleons in the Fermi sea, $\Sigma_{N_{i}}^{2}(k)$. (c) the one-loop diagram in which one nucleon is in the Fermi sea while the other is in vacuum, $\Sigma_{N_{i}}^{3}(k) $. (d)  the double scattering term represented by the 2-loop diagram with two Fermi sea insertions and 1 pion propagator.}
    \label{sigma}
  \end{figure}

Next we consider $ \Sigma^2_{N_i} (k)$ which arises in a diagram (b) in Fig.~\ref{sigma}
\begin{eqnarray*}
\lefteqn{
\Sigma_{N_i}^2 (k)  
} &&  \\
&=& - \int  \frac{d^4 p}{(2 \pi )^4} {\rm Tr_s} \Big{[} ( \slash{k} \gamma_5 ) iD^{-1}_m (p+k) (\slash{k} \gamma_5 )  iD^{-1}_m (p) \Big{]}  \\
&=& \int \frac{d^3 p}{(2 \pi )^3 2E({\bf p})}(-2\pi)  \frac{\delta(k_{0} + E({\bf p}) - E({\bf k+p}))}{2 E({\bf k+p})}  \\
&& \times{\rm Tr}_s \Big{[} \slash{k} (\slash{p} + \slash{k} - m_N ) \slash{k}  (\slash{p} + m_N ) \Big{]} 2 \delta^{ij} \\
&&\times \theta (k_F^i - |{\bf p}+ {\bf k}|) \theta (k_F^i - |{\bf p}|) ,
\end{eqnarray*}
where we have used $\delta ((p+k)^2 - m_N^2) \theta (p^0 + k^0) = \delta(k_{0}+p_{0} - E({\bf k+p}))/(2 E({\bf k+p}))$ with $p_{0}= E({\bf p})$.
Since both nucleons are in the Fermi sea, they are on the mass shell, $p^{2}=m_{N}^{2}$ and $(k+p)^{2}=m_{N}^{2}$, which provides $2k\cdot p + k^{2}=0$. Using these facts, the trace can be evaluated as
\begin{equation}
{\rm Tr_s} \Big{[} \slash{k} (\slash{p} + \slash{k} - m_N ) \slash{k}  (\slash{p} + m_N ) \Big{]}
= - 8 k^{2} m_{N}^{2} .
\end{equation}
Now we consider the $1/m_{N}$ expansion in the nucleon energy and take the first 
term, that is, $E({\bf p}) = E({\bf p+k}) = m_{N}$. Then, finally we obtain
\begin{eqnarray}
\Sigma_{N_i}^2 (k)  &=& \frac{k^{2}}{2\pi^{2}} \delta(k_{0}) 
   \int d^3 p \theta (k_F^i - |{\bf p}+ {\bf k}|) \theta (k_F^i - |{\bf p}|) \nonumber \\
   &=& \frac{k^{2}}{3\pi} \delta(k_{0}) k_F^{i3}
   (1-x)^{2}(x+2) \theta(1-x),  \label{eq:fermiloop}
\end{eqnarray}
where $x=|{\bf k}|/(2k_F^i)$ and we have used the formula \cite{FW}
\begin{eqnarray}
\lefteqn{ \int d^3 p  \, \theta (k_F^i - |{\bf p}+ {\bf k}|) \theta (k_F^i - |{\bf p}|) }
 \nonumber \\ &=& 
  \frac{2\pi}{3} k_F^{i3} (1-x)^{2}(x+2) \theta(1-x). 
\end{eqnarray}

We calculate $ \Sigma^3_{N_i} (k)$ which appears in a diagram (c) shown in Fig.~\ref{sigma} in the soft limit:
\begin{eqnarray}
\Sigma_{N_i}^3 (0)  
&=& \lim_{k\to 0}(-1) \int  \frac{d^4 p}{(2 \pi )^4} \nonumber \\ && 
{\rm Tr}_s \Big{[} ( i \slash{k} \gamma_5 ) iD^{-1}_0 (p+k) ( -i \slash{k} \gamma_5  )  iD^{-1}_m (p) \Big{]} \nonumber \\
&=& \lim_{k\to 0}\int \frac{d^3 p}{(2 \pi )^3 2E({\bf p})} \frac{i}{(k+p)^{2} - m_{N}^{2} +  i \epsilon}  \nonumber \\
&& \times{\rm Tr_s} \Big{[} \slash{k} (\slash{p} + \slash{k} - m_N ) \slash{k}  (\slash{p} + m_N ) \Big{]}  \theta (k_F^i - |{\bf p}|) \nonumber \\
&=& \lim_{k\to 0}\int \frac{d^3 p}{(2 \pi )^3 2E({\bf p})} \frac{i}{k^{2} + 2 k \cdot p+  i \epsilon} \theta (k_F^i - |{\bf p}|) \nonumber \\
&& \times 4  \left[(2 k\cdot p+k^{2}) k\cdot p -  2 k^{2} m_{N}^{2}\right]  \nonumber \\
&=& 0.
\end{eqnarray}
This goes to zero in the soft limit. 

Finally we calculate $ \Sigma^4_{N_i} (k)$ given as diagram (d) in Fig.~\ref{sigma} in the soft limit:
\begin{eqnarray*} 
\lefteqn{
\Sigma_{N_i}^4 (0) } && \\
&=& (-1) \lim_{k \to 0} \int \frac{d^4 p}{(2 \pi )^4} \frac{d^4 q}{(2 \pi )^4} {\rm Tr}_s \Big{[} iD^{-1}_m (q- \frac{p}{2} ) \\
&& \times iD^{-1}_m (q+ \frac{p}{2}) \Big{]} iD_\pi (p+k)  \\
&= & \int \frac{d^3 p}{(2 \pi )^3} \frac{d^3 q}{(2 \pi )^3}
{\rm Tr}_s \Big{[} (\slash{q} - \frac{\slash{p}}{2} + m_N ) (\slash{q} + \frac{\slash{p}}{2} + m_N ) \Big{]} \\
&& \times \frac{ \theta (k_F^i - |{\bf q} - \frac{{\bf p}}{2}|)}{2E({\bf q} - \frac{\bf p}{2})} \frac{\theta (k_F^i - |{\bf q} +  \frac{{\bf p}}{2} |) }{2E({\bf q} + \frac{\bf p}{2})} iD_\pi (p) .
\end{eqnarray*}
Here we have performed the $p_{0}$ and $q_{0}$ integrals, which contain two 
delta functions for the energy conservation and give us the following relation
\begin{eqnarray} 
q_0 &=& \frac{1}{2}( E({\bf q}+\frac{\bf p}{2}) + E({\bf q}-\frac{\bf p}{2}))   \\
p_0 &=&  E({\bf q}+\frac{\bf p}{2}) - E({\bf q}-\frac{\bf p}{2}) 
\end{eqnarray}
In addition we have the following kinematical relations:
\begin{eqnarray}
q \cdot p &=& 0 \\
q^2 + \frac{p^2}{4}  &=& m_N^2 .
\end{eqnarray}
Using these relation, the trace can be evaluated as
\begin{equation}
{\rm Tr}_s \left[ (\slash{q} - \frac{\slash{p}}{2} + m_N ) (\slash{q} + \frac{\slash{p}}{2} + m_N ) \right]
= 8 m_{N}^{2} - 2 p^2 .
\end{equation}
Taking the leading term of the $1/m_{N}$ expansion in which $p_{0}=0$ and $q_{0}=m_{N}$, we obtain  
$\Sigma^4_{N_i}$ as 
\begin{eqnarray} 
\Sigma^4_{N_i} (0) &= & -2i \int \frac{d^3 p}{(2 \pi )^3} \frac{d^3 q}{(2 \pi )^3} \frac{1}{{\bf p}^2 + m_\pi^2} \nonumber \\
&& \times \theta (k_F^i - |{\bf q} - \frac{{\bf p}}{2}|) \theta (k_F^i - |{\bf q} +  \frac{{\bf p}}{2} |) \nonumber \\
&=& -2i G( \frac{m_\pi^2}{4k_F^{i2}}) .
\end{eqnarray}
Here $ G(a) $ is defined by
\begin{displaymath} 
G(a^2) = \frac{k_F^{i4}}{6 \pi^4} \Big{[} \frac{3}{8} - \frac{a^2}{4} -a \arctan \frac{1}{a} + \frac{a^2}{4} (a^2 +3 ) \ln |\frac{1+a^2}{a^2}| \Big{]}.
\end{displaymath}


\begin{thebibliography}{99}
\bibitem{Su}
K. Suzuki et al., Phys. Rev. Lett. {\bf 92}, 072302 (2004)

\bibitem{Friedman:2004jh}
E. Friedman {\it et al.},
Phys. Rev. Lett. {\bf 93}, 122302 (2004).

\bibitem{Friedman:2005pt}
E. Friedman {\it et al.},
Phys. Rev. C {\bf 72},  034609 (2005).

\bibitem{KKW}
E.E. Kolomeitsev , N. Kaiser , and W. Weise , Phys. Rev. Lett. {\bf 90}, 092501 (2003).

\bibitem{JHK}
D. Jido, T. Hatsuda and T. Kunihiro, Phys. Lett. B {\bf 670},  109 (2008).

\bibitem{Hatsuda:1999kd}
T. Hatsuda, T. Kunihiro, and H. Shimizu,
Phys. Rev. Lett. {\bf 82},  2840 (1999).

\bibitem{Jido:2000bw}
D. Jido, T. Hatsuda, and T. Kunihiro,
Phys. Rev. D {\bf 63},  011901 (2000).

\bibitem{Hyodo:2010jp}
T. Hyodo, D. Jido, and T. Kunihiro,
Nucl. Phys. A {\bf 848},  341 (2010).

\bibitem{Weinberg:1967kj}
S. Weinberg,
Phys. Rev. Lett. {\bf 18},  507 (1967).

\bibitem{Kapusta:1993hq}
J.I. Kapusta and E.V. Shuryak,
Phys. Rev. D{\bf 49}, 4694 (1994).

\bibitem{Detar:1988kn}
C.E. DeTar and T. Kunihiro,
Phys. Rev. D{\bf 39},  2805 (1989).

\bibitem{Kim:1998up}
H.C. Kim, D. Jido, and M. Oka,
Nucl. Phys. A {\bf 640} 77 (1998).

\bibitem{Jido:1998av}
D. Jido, Y. Nemoto, M. Oka, and A. Hosaka,
Nucl. Phys. A {\bf 671}, 471 (2000).

\bibitem{Jido:2001nt}
Daisuke Jido, Makoto Oka, and Atsushi Hosaka,
Prog. Theor. Phys. {\bf 106}, 873 (2001).

\bibitem{Jido:2002yb}
D. Jido, H. Nagahiro, and S. Hirenzaki,
Phys. Rev. C {\bf 66}, 045202 (2002).

\bibitem{Nagahiro:2003iv}
H. Nagahiro, D. Jido, and S. Hirenzaki,
Phys. Rev. C {\bf 68},  035205 (2003).

\bibitem{Jido:2008ng}
D. Jido, E. E. Kolomeitsev, H. Nagahiro, and S. Hirenzaki,
Nucl. Phys. A {\bf 811}, 158 (2008).



\bibitem{Lee:1996zy}
S.H. Lee and T. Hatsuda,
Phys. Rev. D {\bf 54} 1871 (1996).

\bibitem{Jido:2011pq}
D. Jido, H. Nagahiro, and S. Hirenzaki,
Phys. Rev. C {\bf 85} 032201(R) (2012).

\bibitem{DL}
E. G. Drukarev and E. M. Levin, Prog. Part. Nucl. Phys. {\bf  27}, 77 (1991).

\bibitem{Coh91}
T.D. Cohen, R.J. Furnstahl, D.K. Griegel, Phys. Rev. C {\bf 45}, 1881 (1992).

\bibitem{KHW}
N. Kaiser, P. de Homont and W. Weise, Phys. Rev. C {\bf 77}, 025204 (2008).


\bibitem{Ikeno:2011mv}
Natsumi Ikeno {\it et al},
Prog. Theor. Phys. {\bf 126} 483  (2011).


\bibitem{Guo94}
Guo-Qiang Li, C.M. Ko,
Phys. Lett. B{\bf 338}, 118 (1994).


\bibitem{Bro96}
R. Brockmann, W. Weise,
Phys. Lett. B{\bf 367} (1996) 40.


\bibitem{Kaiser:2001bx}
N. Kaiser and W. Weise,
Phys. Lett. B{\bf 512}, 283-289 (2001).


\bibitem{Kaiser:2001jx}
N. Kaiser, S. Fritsch, and W. Weise,
Nucl. Phys. A{\bf 697}, 255 (2002).


\bibitem{Girlanda:2003cq}
L. Girlanda, A. Rusetsky, and W. Weise,
Annals Phys. {\bf 312}, 92 (2004).



\bibitem{Doring:2007qi}
M. Doring and E. Oset,
Phys. Rev. C {\bf 77}, 024602 (2008).



\bibitem{O}
J. A. Oller, Phys. Rev. C {\bf 65}, 025204 (2002).
\bibitem{MOW}
U. G. Meissner, J. A. Oller and A. Wirzba, Ann. Phys. {\bf 297}, 27 (2002).


\bibitem{W}
S. Weinberg, Physica A{\bf 96}, 327 (1979).
\bibitem{GL1}
J. Gasser and H. Leutwyler,
Ann. Phys. {\bf 158}, 142 (1984).
\bibitem{GL2}
J. Gasser and H. Leutwyler, Nucl. Phys. B {\bf 250}, 465 (1985) .

\bibitem{GSS}
J. Gasser, M. E. Sainio, and A. Svarc, Nucl. Phys. B {\bf 307}, 779 (1988).

\bibitem{GLS}
J. Gasser, H. Leutwyler, and M. E. Sainio, Phys. Lett. B{\bf 253}, 252 (1991).

\bibitem{LOM}
A. Lacour, J. A. Oller, and U.-G. Meissner, Ann. Phys. {\bf 326}, 241 (2011).

\bibitem{Eri66}
M. Ericson and T. E. O. Ericson, Ann. Phys. (N.Y.) {\bf 36},
323 (1966).

\bibitem{Dmitrasinovic:1999pu} 
  V.~Dmitrasinovic,
  Phys.\ Rev.\ C {\bf 59}, 2801 (1999).

\bibitem{Weinberg:1968de}
S.~Weinberg,
Phys. Rev. {\bf 166}, 1568 (1968).

\bibitem{Charap:1971bn}
J.M. Charap,
Phys. Rev. D {\bf 3}, 1998 (1971).

\bibitem{Gerstein:1971fm}
I.S. Gerstein, R. Jackiw, S. Weinberg, and B.W. Lee,
Phys.Rev. D {\bf 3}, 2486 (1971).

\bibitem{FW}
A.L. Fetter and J.D. Walecka,
{\it Quntaum theory of many-particle systems}
(Dover, New York, 2003)

\end{thebibliography}
\end{document}